\documentclass[10pt]{article}

\usepackage{amsmath}
\usepackage{amssymb}

\usepackage{graphicx}

\usepackage{cite}

\usepackage{color} 

\usepackage[labelfont=bf,labelsep=period,justification=raggedright]{caption}

\makeatletter
\renewcommand{\@biblabel}[1]{\quad#1.}
\makeatother

\date{}

\renewcommand\Re{\operatorname{Re}}
\renewcommand\Im{\operatorname{Im}}

\usepackage{epstopdf}

\begin{document}

% Title must be 150 characters or less
\begin{flushleft}
{\Large
\textbf{Cross-correlation asymmetries and causal relationships between stock and market risk}
}
% Insert Author names, affiliations and corresponding author email.
\\
Stanislav S. Borysov$^{1,2,3,\ast}$, 
Alexander V. Balatsky$^{1,4}$
\\
\bf{1} Nordita, KTH Royal Institute of Technology and Stockholm University, Roslagstullsbacken 23, SE-106 91 Stockholm, Sweden
\\
\bf{2} Nanostructure Physics, KTH Royal Institute of Technology, Roslagstullsbacken 21, SE-106 91 Stockholm, Sweden
\\
\bf{3} Theoretical Division, Los Alamos National Laboratory, Los Alamos, NM 87545, USA
\\
\bf{4} Institute for Materials Science, Los Alamos National Laboratory, Los Alamos, NM 87545, USA
\\
$\ast$ E-mail: borysov@kth.se
\end{flushleft}

% Please keep the abstract between 250 and 300 words
\section*{Abstract}
We study historical correlations and lead-lag relationships between individual stock risk (volatility of daily stock returns) and market risk (volatility of daily returns of a market-representative portfolio) in the US stock market. We consider the cross-correlation functions averaged over all stocks, using 71 stock prices from the Standard \& Poor's 500 index for 1994--2013. We focus on the behavior of the cross-correlations at the times of financial crises with significant jumps of market volatility. The observed historical dynamics showed that the dependence between the risks was almost linear during the US stock market downturn of 2002 and after the US housing bubble in 2007, remaining on that level until 2013. Moreover, the averaged cross-correlation function often had an asymmetric shape with respect to zero lag in the periods of high correlation. We develop the analysis by the application of the linear response formalism to study underlying causal relations. The calculated response functions suggest the presence of characteristic regimes near financial crashes, when the volatility of an individual stock follows the market volatility and vice versa.
% Please keep the Author Summary between 150 and 200 words
% Use first person. PLoS ONE authors please skip this step. 
% Author Summary not valid for PLoS ONE submissions.   
%\section*{Author Summary}

\section*{Introduction}

A financial market is a complex system demonstrating diverse phenomena and attracting attention from a whole spectrum of disciplines ranging from social to natural science\cite{econPhysNext}. Better understanding of the behavior of financial markets has become an integral part of the discussion on further sustainable economic development. In this context, proper assessment of financial risks \cite{systemic_risk_survey} plays a crucial role: Underestimated risks contribute to financial bubbles with eventual crashes while overestimation of risks might cause inefficiency of financial resource allocations and a slowdown in economic growth, giving rise to periods of stagnation. This multifaceted problem, lying at the core of finance, draws significant interest from the physical and mathematical communities\cite{Econophysics, Bouchaud}. One of the key components of financial risk analysis is a volatility assessment, which quantifies the financial stability of an asset in question. To this end, a number of methods have been proposed for risk modeling \cite{LeBaron19991487, vol_model, vol_multi, Miccich2002756} and forecasting\cite{vol_review}, along with numerous studies of various empirical properties of volatility, including such stylized facts as clustering\cite{PhysRevE.60.1390, PhysRevE.60.5305, PhysRevLett.89.158701}, lead-lag effects\cite{Lo01041990}, asymmetries \cite{Bekaert01012000,asymmetry1} and many others (for a review see Refs.~\cite{doi:10.1080/14697688.2010.539248, doi:10.1080/07474930701853509}). Related phenomena, being a result of collective behavior, also involve such aspects as estimation of correlation \cite{Erb1994,PhysRevLett.83.1467,asymmetr_corr_mat} and cross-correlation\cite{PhysRevE.65.066126, PhysRevLett.83.1471, PhysRevE.70.026110, PhysRevE.63.061101} matrices, study of their dynamics\cite{PhysRevE.84.026109,0295-5075-90-6-68001}, asymmetric correlations \cite{JOFI:JOFI340}, nonlinear correlations\cite{e15114909,copulas,lee2000statistics} and detrending \cite{szekely2009, PhysRevLett.100.084102}, financial networks and clustering\cite{dx.doi.org/10.1007/s100510050929, Schweitzer24072009, doi:10.1080/14697680400020325, PhysRevE.68.046130, PhysRevE.72.046133, Heimo2009145, e2009-00419-5, e2009-00255-7, Coelho2007615, 10.1371/journal.pone.0015032}, multivariate stochastic models\cite{doi:10.1080/07474930600713564, PhysRevE.75.016106}, critical phenomena \cite{PhysRevE.88.062814, Majdandzic2014}, etc.

\begin{figure}[!ht]
\begin{center}
\includegraphics[width=3in]{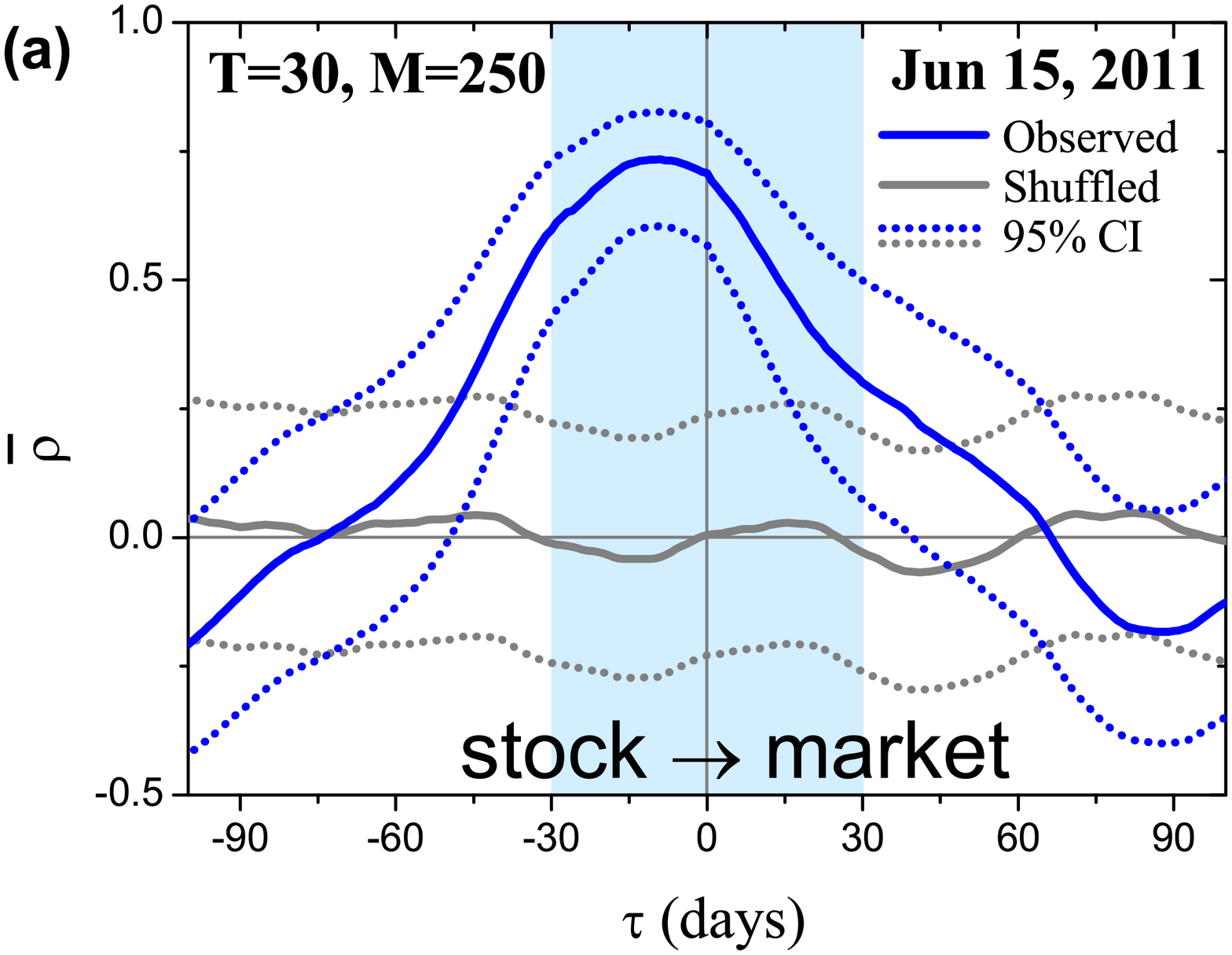}
\hspace{5mm}
\includegraphics[width=3in]{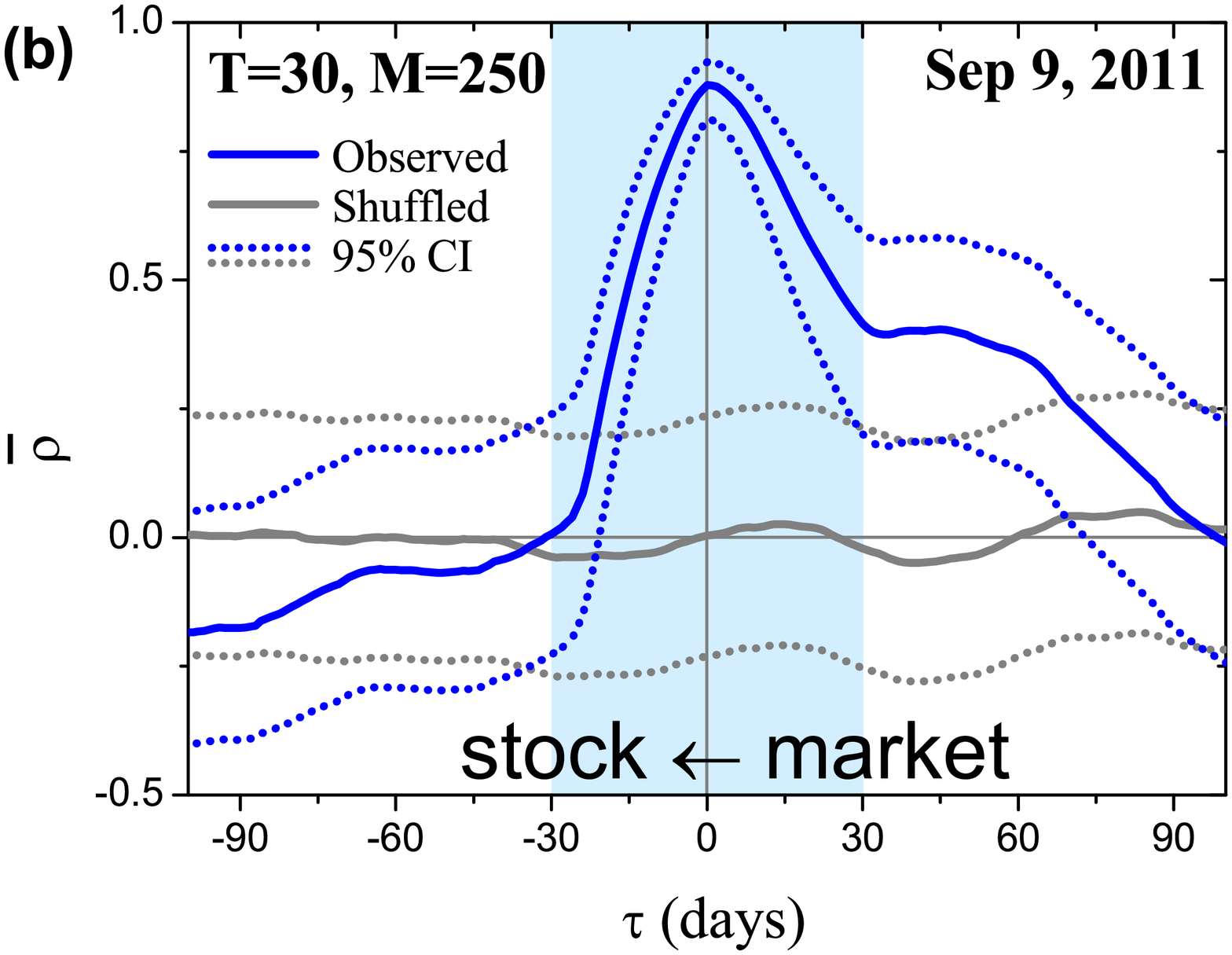}
\end{center}
 \caption{
{\bf Cross-correlation function $\overline{\rho}(\tau)$ between stock and market volatility (blue solid line) averaged over $N = 71$ stocks for two different dates (Jun 15, 2011 and Sep 9, 2011) near the European sovereign debt crisis.} The cross-correlations possess asymmetry with respect to zero lag ($\tau=0$): (a) changes in individual stock risks on average precede changes in the market risk with lag of 14 days; (b) individual stock risks on average are prone to follow the market risk. Stock and market volatilities are calculated using an SMA with the window $T=30$ days. The cross-correlations between them are calculated using an SMA with the window $M=250$ days. Highlighted ranges with a blue background around zero lag ($\pm 30$ days) are further used for the calculation of the susceptibilities depicted in Fig.~\ref{fig:vol_vs_vol_m_suscept}. Grey solid line corresponds to $\overline{\rho}(\tau)$ when the underlying stock returns are randomly shuffled. The corresponding 95\% confidence intervals for the mean correlations are denoted with dotted lines.
}
\label{fig:vol_vs_vol_m_cross_corr}
\end{figure}

In the current paper, we focus on lead-lag effects between individual and collective volatility behavior in the US stock market, which might be further discussed in the context of the systemic regulation problem \cite{Beale18072011}. Former studies reported an increase of correlations across financial markets in recent times \cite{PhysRevE.84.026109} along with overall market disposition to systemic collapses \cite{10.1371/journal.pone.0019378}. Our investigation thus has an aim to shed additional light on the dynamics of systemic risk in the last decade. For this purpose, we analyze historical prices of 71 stocks (Table~\ref{tab:companies}) from the Standard \& Poor's 500 index \cite{sp500} (hereafter S\&P 500) for 1994--2013. Although we employ one of the simplest volatility estimators---the simple moving average (SMA) standard deviation of daily logarithmic returns---it is conjectured to correctly describe asset risk dynamics on long time scales, on the order of months and years \cite{handbook_finance}. We harness cross-correlation analysis which is a basic tool in the analysis of multiple time series. By definition, the absolute value of the normalized cross-correlation function lies between 0 and 1, indicating the strength of a linear relationship between time series, given that one is shifted by a particular lag value. It is crucial to note that our approach is based on a study of cross-correlations between {\it derived} quantities from the stock returns (standard deviations) rather than the analysis of cross-correlation matrices of the returns {\it per se}, implicitly involving calculation of cross-correlations between correlations\footnote{Since portfolio return is the sum of stock returns, its variance is the sum of all elements of the covariance matrix, $\mathbf{C}$, [Eq.~(\ref{eq:market_vol})]. On the other hand, any covariance matrix can be factorized into the product of a correlation matrix, $\mathbf{R}$, and a diagonal matrix of standard deviations, $\mathbf{C}^\mathrm{diag}$, with elements $C^\mathrm{diag}_{ii} = \sqrt{C_{ii}}$ and $C^\mathrm{diag}_{i\neq j}=0$: $\mathbf{C} = \mathbf{C}^\mathrm{diag} \mathbf{R} \mathbf{C}^\mathrm{diag}$.}. These more sophisticated quantities will hopefully allow us to capture a more systematic evolution of the market risk as a function of time. Indeed, it was previously shown that market volatility and correlation are tightly related across international financial markets \cite{Solnik1996}. However, our calculations show that the cross-correlation function averaged over all stocks (see equations below) not only often has the maximum value close to 1 but also possesses an asymmetric shape with respect to zero lag (Fig.~\ref{fig:vol_vs_vol_m_cross_corr}). These features suggest the presence of long-term trends, when equilibrium on the market is not reached within one trading day and overall market risk tends to follow individual stock risks [Fig.~\ref{fig:vol_vs_vol_m_cross_corr}(a)] or vice versa [Fig.~\ref{fig:vol_vs_vol_m_cross_corr}(b)]. Lately, emergence of intraday trends has been reported for stock returns \cite{2014arXiv1401.0462C} and correlations \cite{0295-5075-99-3-38001}, while our investigation develops similar ideas for stock volatilities.

Generally, it is not possible to determine causality from an arbitrary shape of the cross-correlation function. However, if the cross-correlation function is asymmetric with respect to the time reversal operation (change of a sign of the time lag), it might hint at the presence of causal relationships\cite{Pierce1977265}. Although determining {\it true} causality is rather a philosophical matter, we use this term in the {\it predictive} sense, i.e. if the past values of one time series can be used to predict the present or future values of the other. In this regard, one of the most widely used approaches is the Granger causality test \cite{Granger1969}. Following this method, one builds autoregressive models for the time series including and excluding factors in question and checks if the difference between models is statistically significant. However, in the current investigation, we propose to use an alternative approach utilizing a specific class of asymmetric cross-correlation functions studied in linear response theory \cite{JPSJ.12.570}, which provides a framework for describing input-output properties of a physical system. Within this approach, causality implies the absence of any response before an action (as long as there are no long-term memory effects), that results in zero values of the cross-correlation function for a particular lag direction---positive or negative---depending on the input-output roles of the variables. The simplest example can be given by a force acting on a mass. The mass cannot move before the interaction and thus the correlation between the force and displacement is zero before the time when the force is applied. Although we do not expect to observe such a trivial behavior in real financial markets, asymmetries in the empirical functions (Fig.~\ref{fig:vol_vs_vol_m_cross_corr}) can be interpreted as an approximation to this ideal model, where the mass and force are represented by individual stock and collective market volatility or vice versa, depending on the observed regime. Making use of this approximation, we restrict ourselves to the qualitative analysis with aim to reveal historical patterns only.

%-------------------------------------------------------------------------------
\section*{Estimating the stock and market risks}
Let us first introduce notations used throughout the paper. We consider $N$ discrete time series of daily closing stock prices $S_{i}(t)$, $i=1,\dots,N$ which are converted to log-returns $s_{i}(t) = \ln[S_{i}(t)/S_{i}(t-1)]$, assuming continuous compound interest. Within the SMA approach, one can calculate a moving average for a particular discrete time moment $t$ using equally weighted values of $T$ previous days including the current one
\begin{equation}
	\langle s_{i}(t) \rangle := \frac{1}{T}\sum\limits_{t'=t-T+1}^{t}s_{i}(t').
	\label{eq:average}
\end{equation}
In this case, a cross-covariance of two time series might be defined as
\begin{equation}
	\sigma[s_i,s_j](t,\tau) := \langle s_{i}(t+\tau)s_{j}(t) \rangle - \langle s_{i}(t+\tau)\rangle \langle s_{j}(t) \rangle,
	\label{eq:cov}
\end{equation}
where $\tau$ is a time lag. Series variance is a self-covariance at $\tau=0$, $\sigma^2[s_i](t) \equiv \sigma[s_i,s_i](t,0)$, where $\sigma$ denotes the standard deviation or volatility in finance. This quantity can be used as the simplest risk measure: Stocks with higher values of $\sigma$ have less stable returns and, consequently, are less attractive for investment, other things being equal.

\begin{figure}[!ht]
\begin{center}
\includegraphics[width=6in]{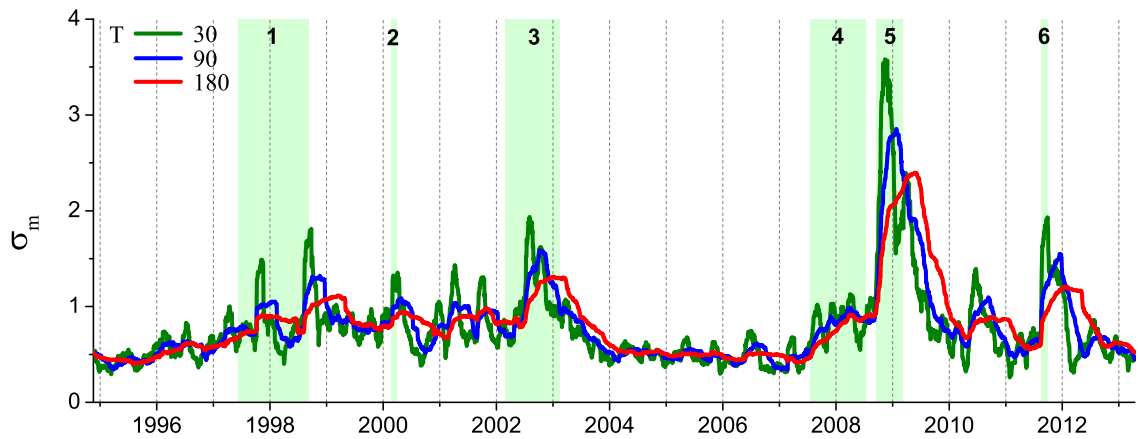}
\end{center}
 \caption{
{\bf Historical dynamics of the US stock market volatility $\sigma_\mathrm{m}$.} It is represented by the SMA standard deviation of returns of the portfolio consisting of $71$ US stocks (Table~\ref{tab:companies}), calculated using windows of $T=30$ (green line), $90$ (blue line) and $180$ (red line) days. The distance between two labeled dates is 500 trading days. Market crashes correspond to abrupt jumps of the volatility. Use of the bigger values of $T$ leads to smoothing of small crashes, while the biggest ones are still clearly seen. Main financial crises are highlighted with a light green background: (1) Asian and Russian crisis of 1997--1998, (2) dot-com bubble, (3) US stock market downturn of 2002, (4) US housing bubble, (5) bankruptcy of Lehman Brothers followed by the global financial crisis, (6) European sovereign debt crisis.
}
\label{fig:crashes}
\end{figure}

A stock market comprises all stocks available for trade. Although in the current investigation we consider a limited subset of stocks, it is chosen to represent the top US companies with the largest market capitalization. For such a portfolio, consisting of equal shares of $N$ stocks, total return, $m(t)$, equals to the sum of the separate stock returns, $m(t) = \sum_{i=1}^N s_{i}(t)$. Its variance, in addition to Eq.~(\ref{eq:cov}), can be also expressed as the sum of all elements of the covariance matrix $\mathbf{C}(t)$, an $N\times N$ matrix with elements $C_{ij}(t) = \sigma[s_i,s_j](t,0)$,
\begin{equation}
	\sigma^2[m](t) = \sum\limits_{i=1}^N\sum\limits_{j=1}^N C_{ij}(t).
	\label{eq:market_vol}
\end{equation}
The square root of this value, $\sigma_\mathrm{m} \equiv \sigma[m]$, can be also used as a portfolio risk measure, which characterizes overall market risk in the case of large $N$ (Fig.~\ref{fig:crashes}). In the remainder of the paper, we will focus on finding historical dependences and lead-lag relationships between individual stock risks, $\sigma_i \equiv \sigma[s_i]$, and market risk, $\sigma_\mathrm{m}$, using the formalism presented in the following section.

%-------------------------------------------------------------------------------
\section*{Causality analysis}
One of the possible ways to estimate dependence between two time series $x(t)$ and $y(t)$ is to calculate the cross-correlation function
\begin{equation}
	\rho[x,y](t,\tau) = \frac{\sigma[x,y](t,\tau)}{\sigma[x](t+\tau)\sigma[y](t)},
	\label{eq:corr}
\end{equation}
which is normalized and ranges from $-1$ to $1$. Its peak value\footnote{In this section, we assume this peak value to be positive, since the opposite case can be easily recovered via multiplication of $x$ or $y$ by $-1$. Noting that $\sigma[-x,y]=\sigma[x,-y]=-\sigma[x,y]$ and $\sigma[-x]=\sigma[x]$, one immediately gets $\rho[-x,y]=\rho[x,-y]=-\rho[x,y]$} shows the strength of a linear relationship between $x$ and $y$ (with zero value corresponding to its absence) when the first series is shifted by the time lag $\tau$. If the dependence between series is nonlinear, more sophisticated statistical concepts should be used instead, for instance, cross-entropy\cite{e15114909}, copula\cite{copulas} or the Spearman's rank correlation\cite{lee2000statistics}. However, we are aimed to employ the linear Pearson's coefficient [Eq.~(\ref{eq:corr})] in the present study. Given two series are correlated, it is not possible to establish causal relationships between the variables by this fact itself. However, the particular shapes of the cross-correlation functions studied within linear response theory can provide an insight into this problem.

\begin{figure}[!ht]
\begin{center}
\includegraphics[width=3in]{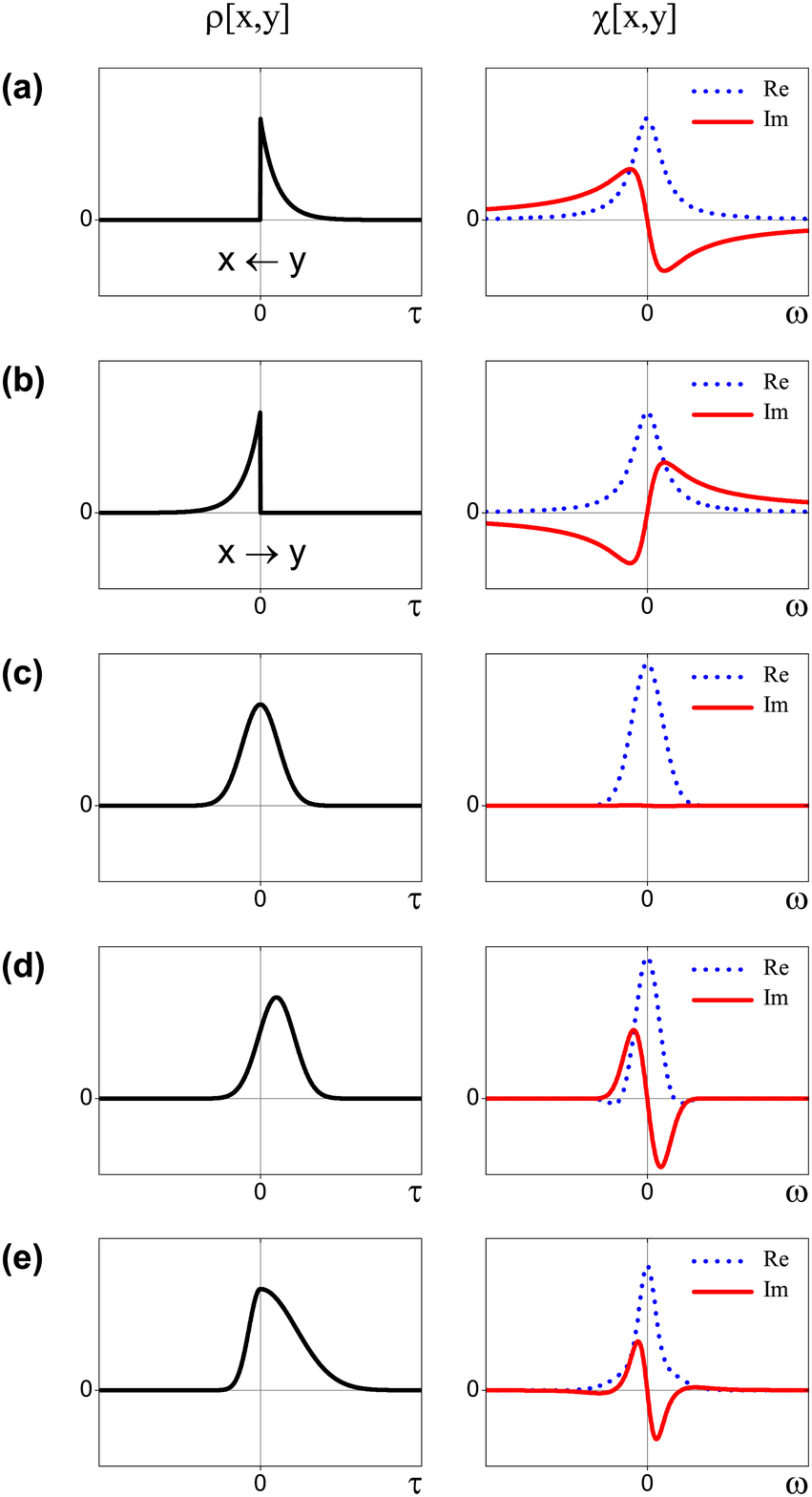}
\end{center}
 \caption{
{\bf Examples of the cross-correlation functions $\rho$ between two time series $x$ and $y$ (left column); their Fourier transforms $\chi$ (right column).} (a) Impulse response function corresponding to the fundamental solution of Eq.~(\ref{eq:ode}); (b) impulse response function of the same equation with the variables $x$ and $y$ being interchanged. Symmetric shape of $\rho$ results in the zero imaginary part of $\chi$ (c), while its small shift (d) results in the qualitatively similar behavior of the imaginary part as for the impulse response. The Fourier transform of the cross-correlation function which decays to zero with different speed for negative and positive lag values (e) also demonstrates the similar features.
}
\label{fig:method}
\end{figure}

This theory provides a convenient framework for the study of related dynamical properties of a physical system. Within this approach, the cross-correlation function defines the system's response to an external action, obeying laws of motion. In this context, causality implies the absence of any deterministic response before an action, i.e. the expected value of the cross-correlation function is zero for a particular lag direction ($\tau>0$ or $\tau<0$) defined by the input-output roles of $x$ and $y$. For example, the response function of the first-order ordinary differential equation
\begin{equation}
	a\dot x+bx=y,
	\label{eq:ode}
\end{equation}
where $a$ and $b$ are some constants and $y$ is the delta function (impulse force), is depicted in Fig.~\ref{fig:method}(a). Here, $y$ can be uniquely identified as an external action because $\rho[x,y]$ is non-zero only for $\tau>0$, the time direction corresponding to the {\it future} values of $x$ and the {\it past} values of $y$ [see Eq.~(\ref{eq:cov})]. This asymmetry of the response is also graphically reflected in its Fourier transform\footnote{Since we work with discrete time series, we use its discrete analogue with a unitary norm $\chi_\omega = 1/\sqrt{\tau^\mathrm{max}}\sum_{\tau=-\tau^\mathrm{max}}^{\tau^\mathrm{max}}\rho_\tau e^{-i2\phi\omega\tau/\tau^\mathrm{max}}$.} known as susceptibility
\begin{equation}
	\chi(\omega) := \int\limits_{-\infty}^{\infty}\rho[x,y](\tau)e^{-i\omega\tau}d\tau,
	\label{eq:susceptibility}
\end{equation}
which is a complex-valued function of frequency $\omega$. Its real ({\it reactive}) part, $\Re\chi$, being an even function of $\omega$, is defined by the correlation strength. While the imaginary ({\it dissipative}) part, $\Im\chi$, is an odd function of $\omega$ defined by the asymmetric part of $\rho$\footnote{Any function $\rho(\tau)$ can be written as the sum of an even function $\rho^\mathrm{even}(-\tau) = \rho^\mathrm{even}(\tau)$ and an odd function $\rho^\mathrm{odd}(-\tau) = -\rho^\mathrm{odd}(\tau)$. In this case, $\Re\chi$ is the Fourier transform of $\rho^\mathrm{even}$ while $\Im\chi$ is the Fourier transform of $\rho^\mathrm{odd}$.}. Regarding the action-reaction roles of $x$ and $y$ in Eq.~\ref{eq:ode}, $\Im\chi$ has a negative peak for $\omega>0$ [Fig.~\ref{fig:method}(a)] and a positive peak if the variables are interchanged [Fig.~\ref{fig:method}(b)]. Additionally, $\Re\chi$ and $\Im\chi$ satisfy the Kramers-Kronig relations, which is a mathematical condition of a complex function to be analytic and hence the underlying physical system to be stable\cite{PhysRev.104.1760}.

The empirical cross-correlation functions (Fig.~\ref{fig:vol_vs_vol_m_cross_corr}), which characteristic shapes are schematically depicted in Figs.~\ref{fig:method}(c)--(e), differ from the ones studied in linear response theory [Figs.~\ref{fig:method}(a)--(b)]. Despite this fact, the corresponding susceptibilities display the similar features of the real and imaginary parts (Fig.~\ref{fig:vol_vs_vol_m_suscept}). Thus, we consider them as a coarse approximation to the theoretical linear response functions and utilize the peak of $\Im\chi(\omega>0)$ as an indicator of possible causal dependence. If the cross-correlation function is completely symmetric with respect to the time reversal operation [Fig.~\ref{fig:method}(c)], $\tau\to-\tau$, no causal relation between $x$ and $y$ can be established within the linear response formalism given the cross-correlation function alone: This fact implies that the interchange of the input-output roles of the underlying variables produces exactly the same observable behavior of the system as a whole. However, when the maximum value of $\rho$ is slightly shifted [Fig.~\ref{fig:method}(d)] or the function decays faster for the one lag direction than for the other [Fig.~\ref{fig:method}(e)] one might expect that the change of $y$ tends to cause the reaction of $x$ because of the enhanced response for the future values of $x$ . In doing so, reversal of the observed input-output roles corresponds to the change of the sign of the imaginary part while the real part remains unaffected.

Finally, fitting of a particular susceptibility model to the empirical data allows one to determine the differential equation which governs the observed behavior of the system. However, the behavior of a real financial market is usually highly nonlinear, possessing long-term memory effects\cite{10.1007/s100510050292, taylor1986modelling} and fractal structure\cite{Fractal1989, PhysRevE.88.062912}, that is obviously beyond the scope of the discussed method. One of the possible ways to extend the presented approach might be the application of nonlinear response theory\cite{nonlinear_response} although this case is not considered in our paper. We restrict ourselves to the linear qualitative analysis which only hints at the direction of influence between the variables in question.

%-------------------------------------------------------------------------------
\section*{Results}

\begin{figure}[!ht]
\begin{center}
\includegraphics[width=2in]{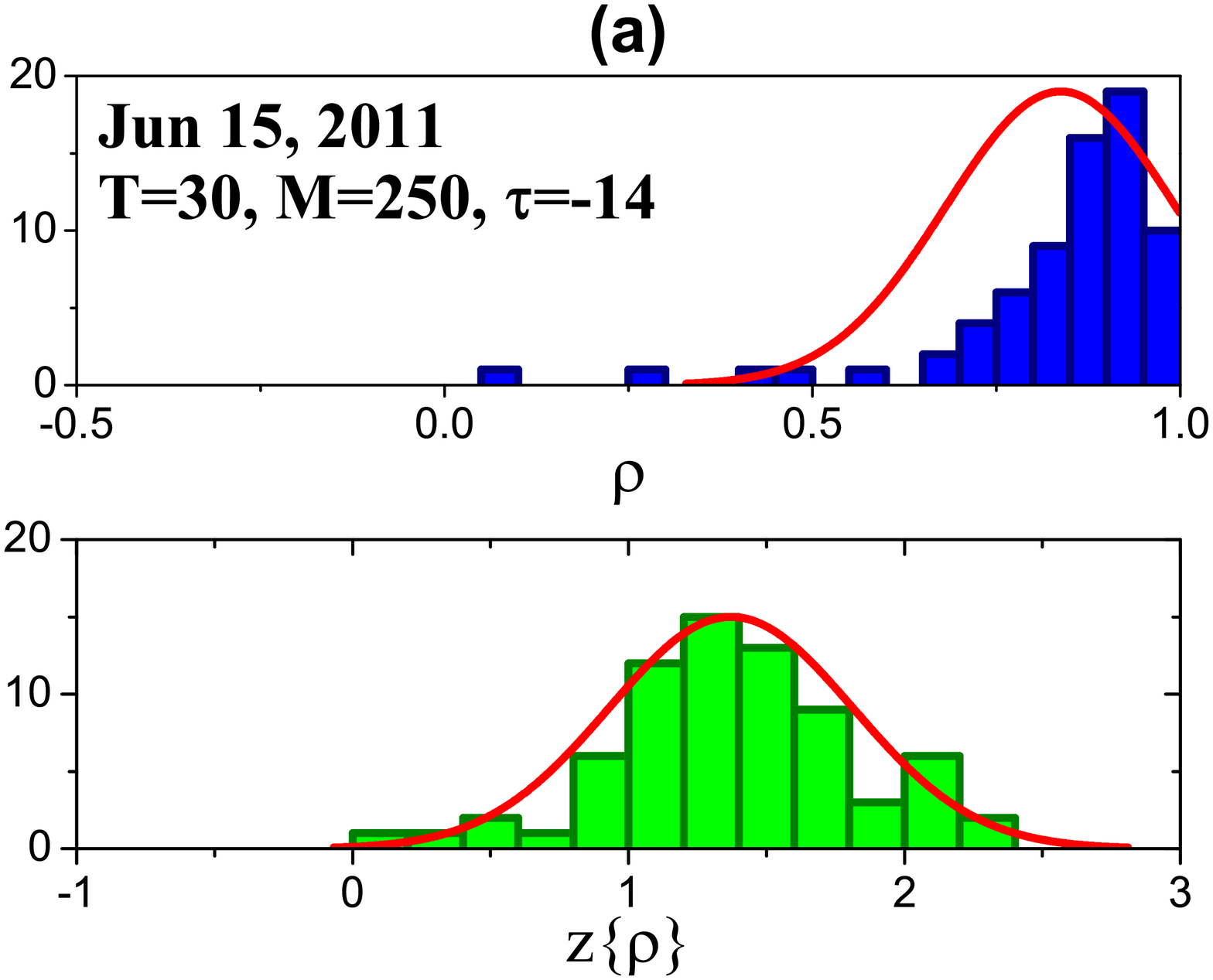}
\hspace{0mm}
\includegraphics[width=2in]{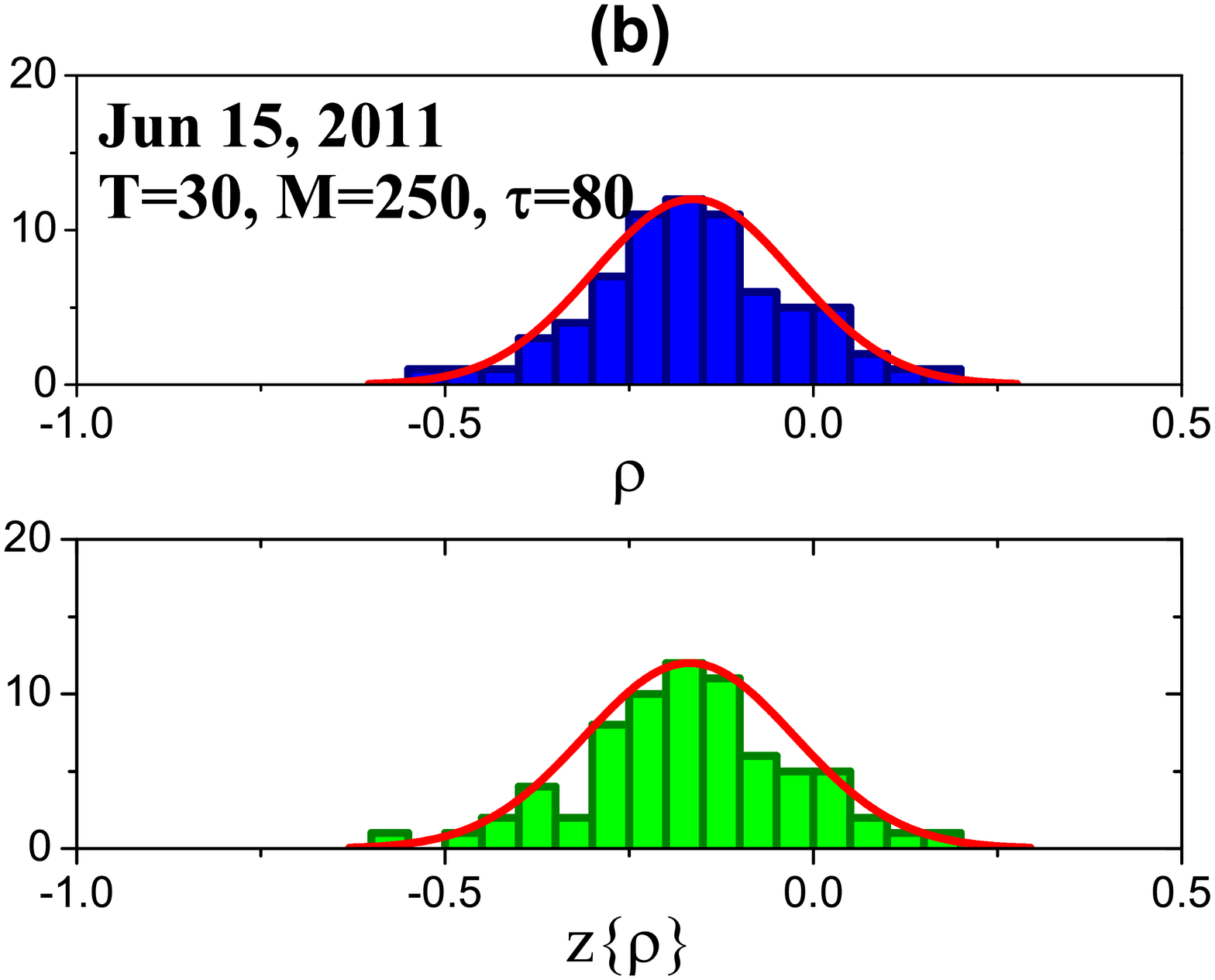}
\hspace{0mm}
\includegraphics[width=2in]{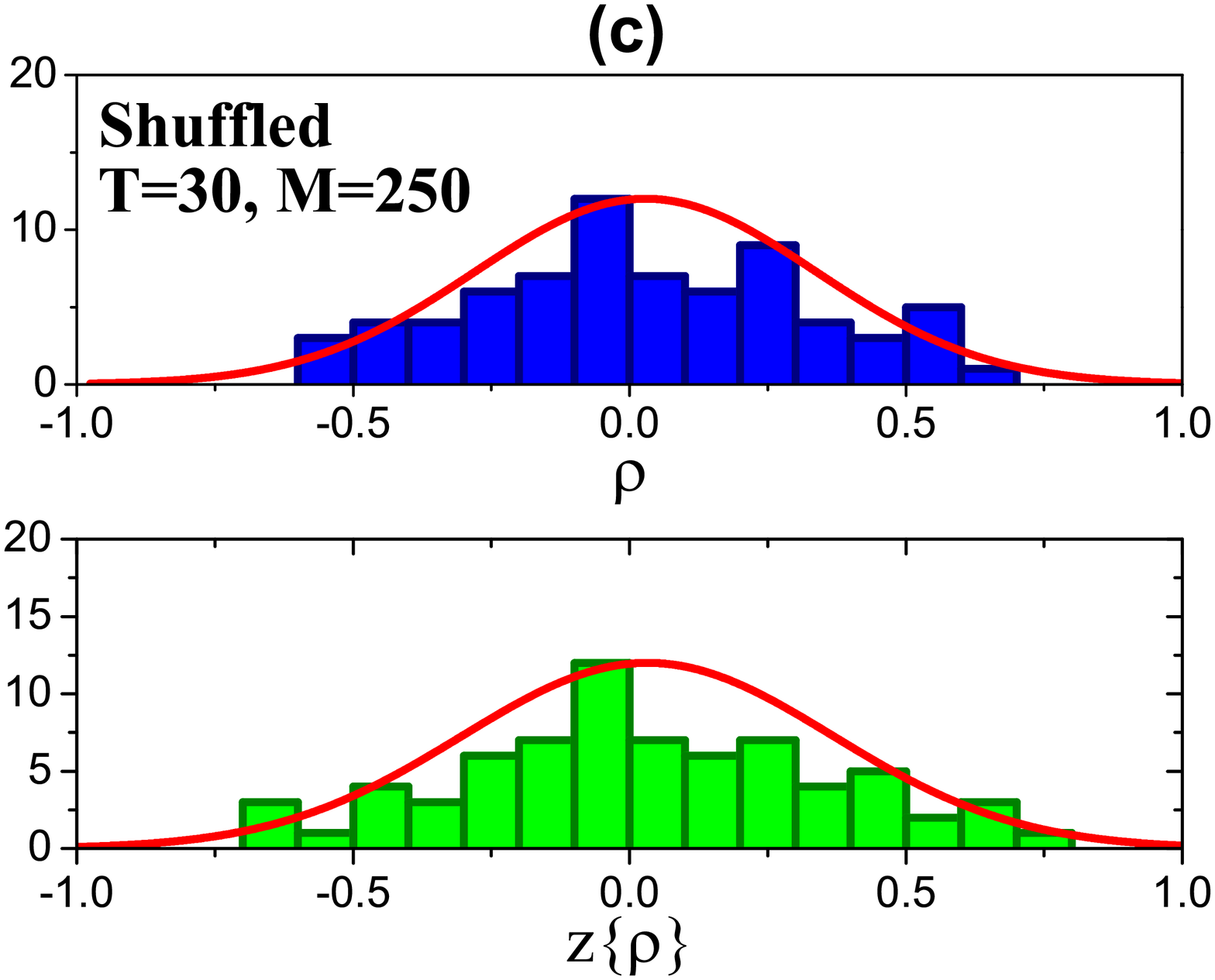}
\end{center}
 \caption{
{\bf Histograms of correlation coefficients (top) and their Fisher transforms (bottom) on Jun 15, 2011 for lags $\tau = -14$ (a) and $80$ (b) days, and randomly shuffled returns (c).} Stock and market volatilities are calculated using an SMA with the window $T=30$ days. The correlations between them are calculated using an SMA with the window $M=250$ days. The red curves denote fitted normal distribution. In the case of large correlations, the Fisher transform makes the highly skewed distribution approximately Gaussian (a). 
}
\label{fig:rho_distrib}
\end{figure}

\begin{figure}[!ht]
\begin{center}
\includegraphics[width=2in]{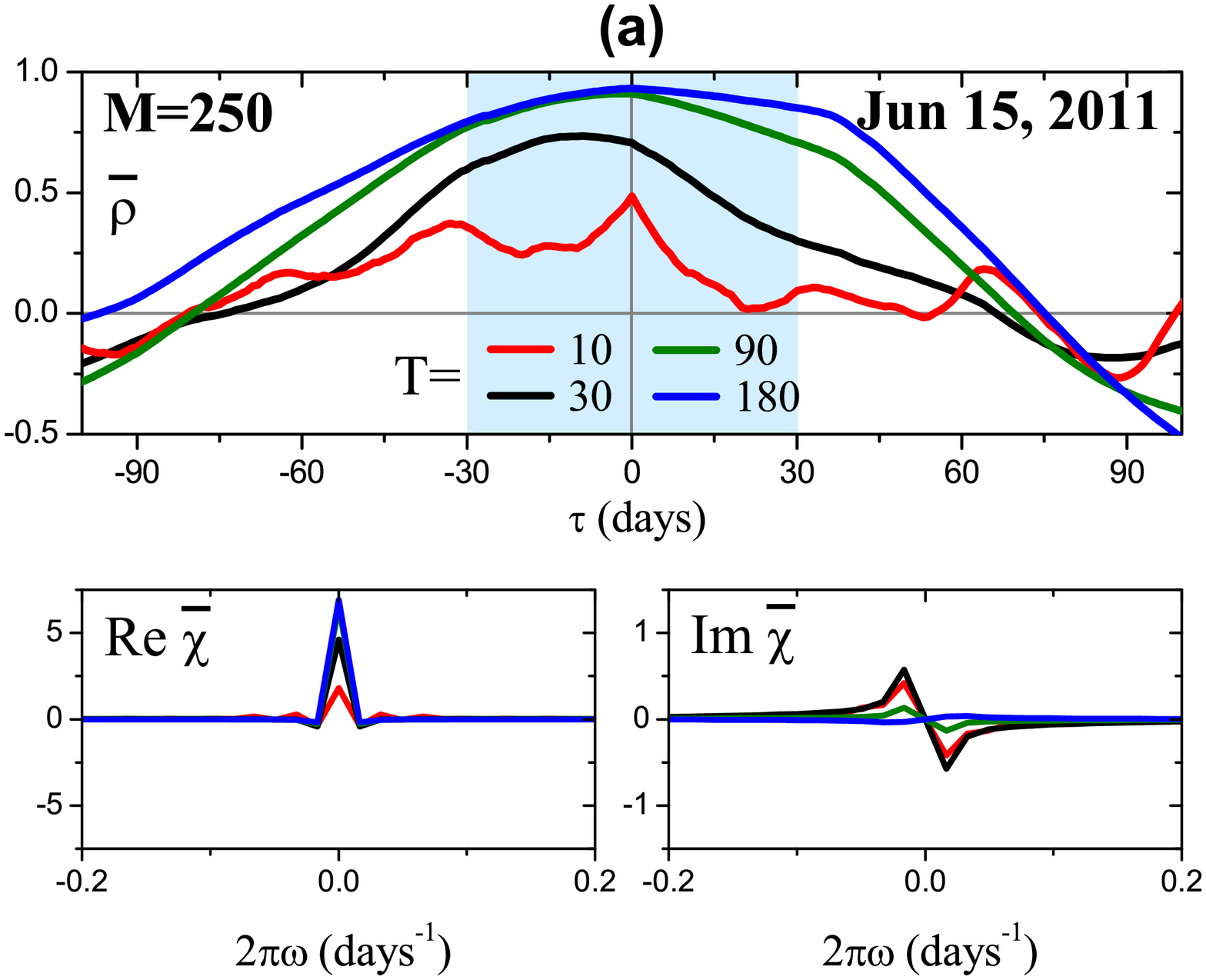}
\hspace{0mm}
\includegraphics[width=2in]{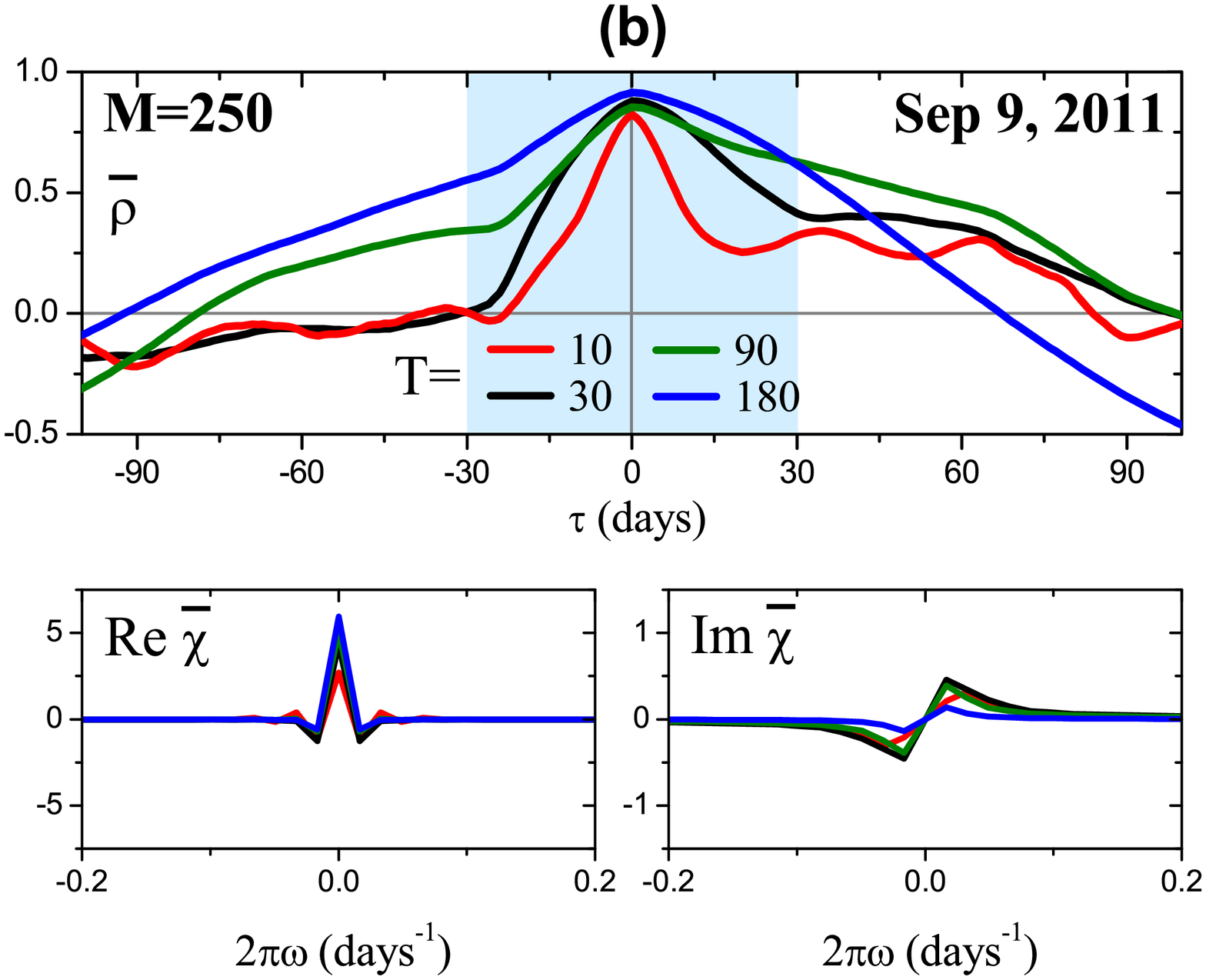}
\hspace{0mm}
\includegraphics[width=2in]{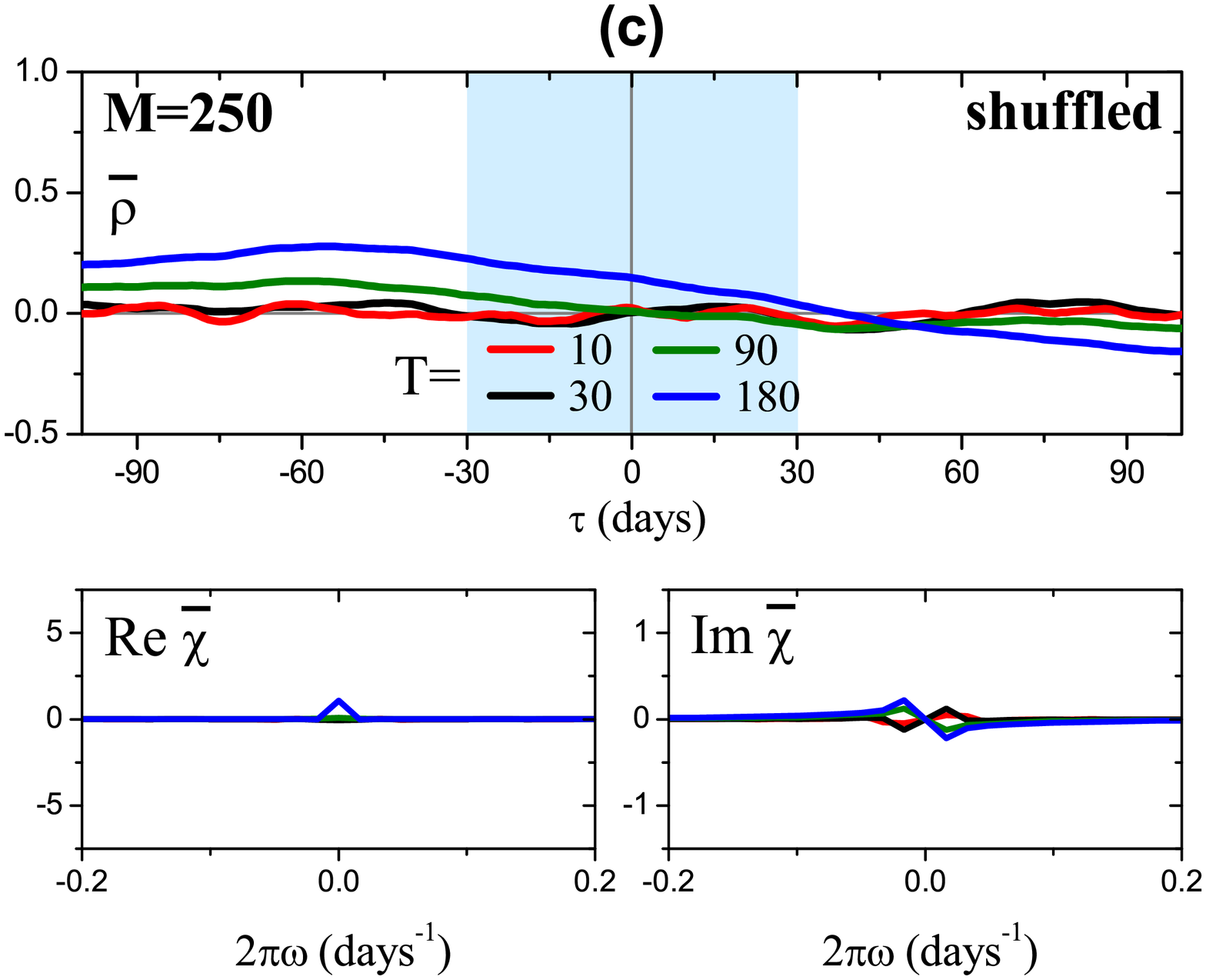}\\
\includegraphics[width=2in]{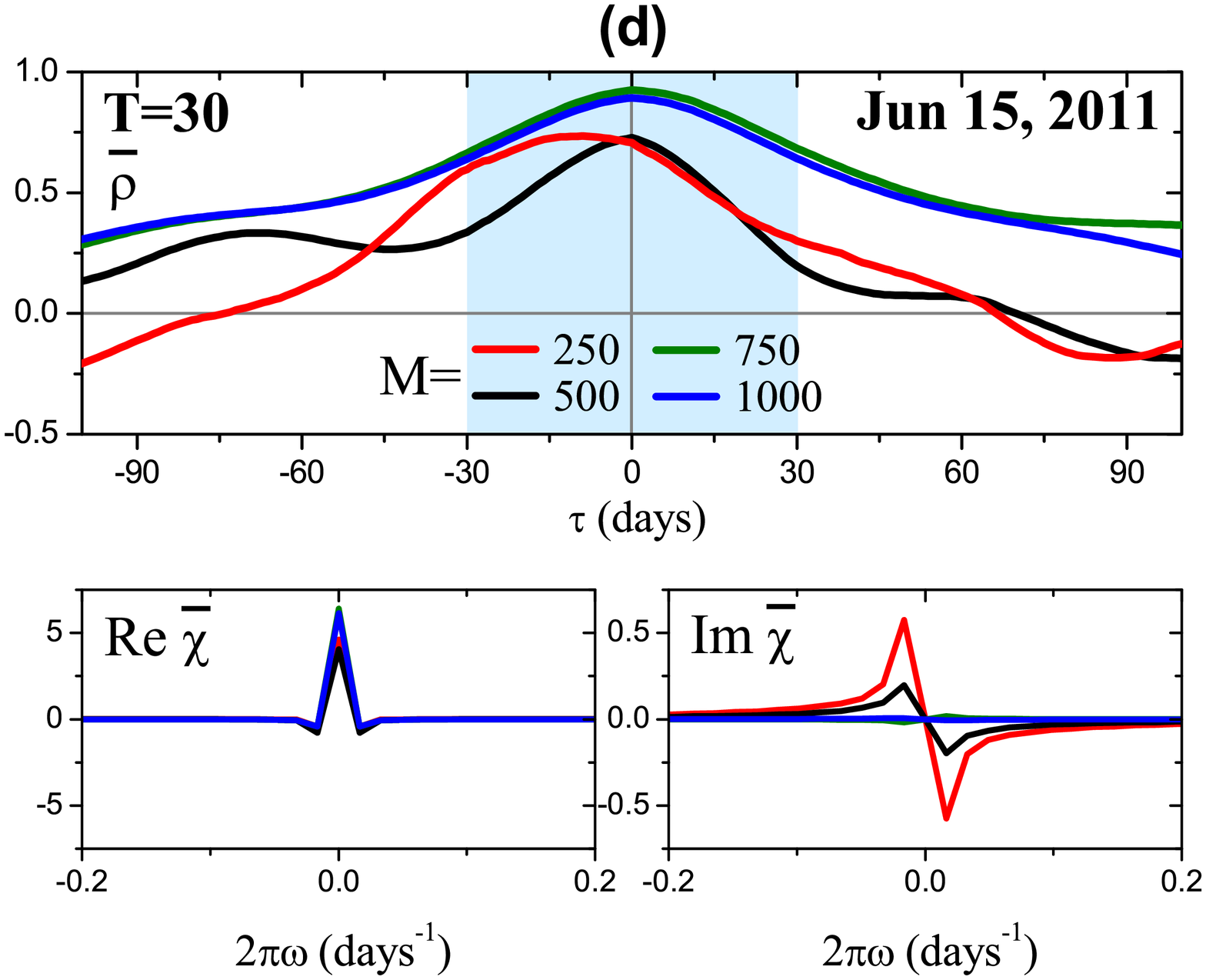}
\hspace{0mm}
\includegraphics[width=2in]{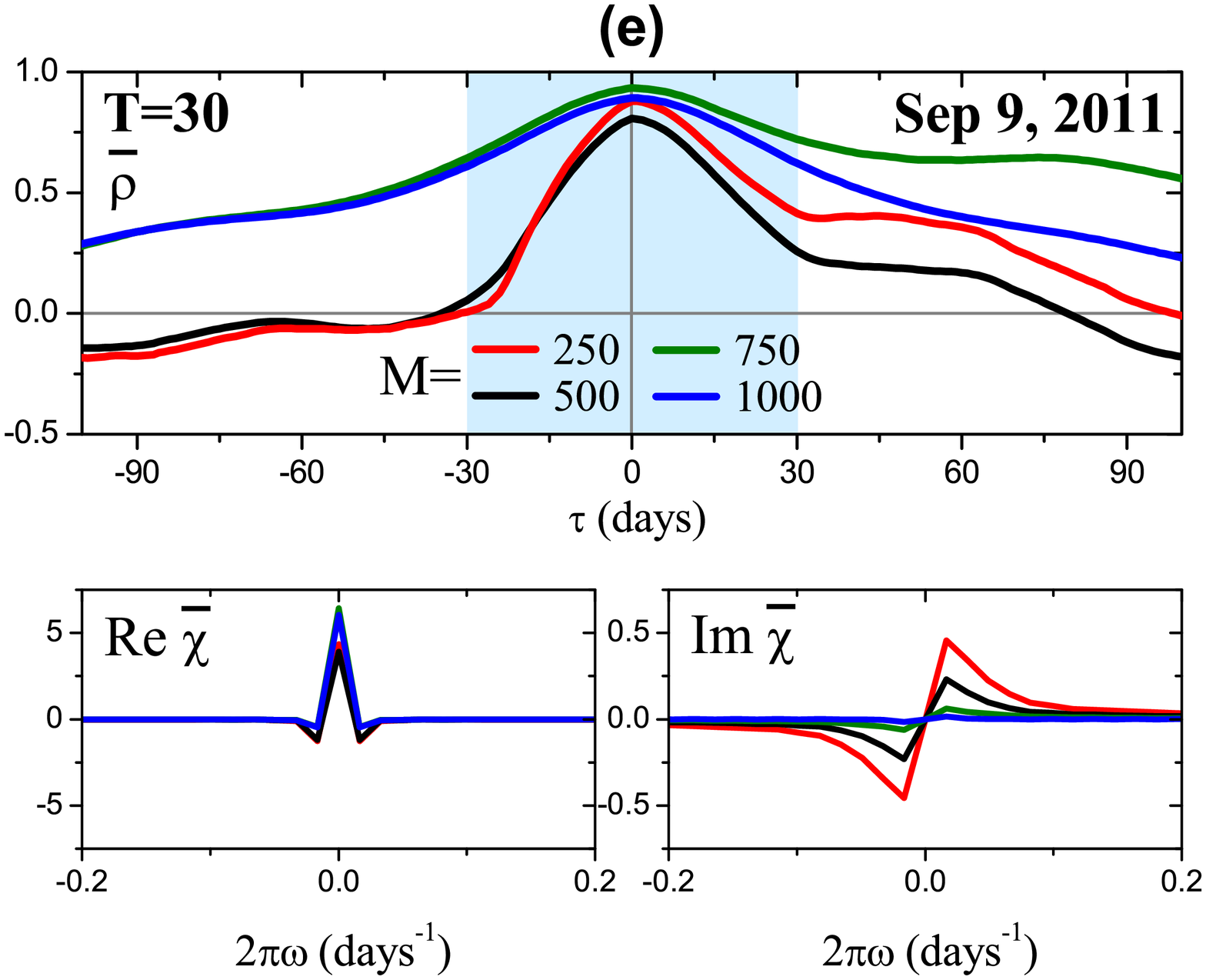}
\hspace{0mm}
\includegraphics[width=2in]{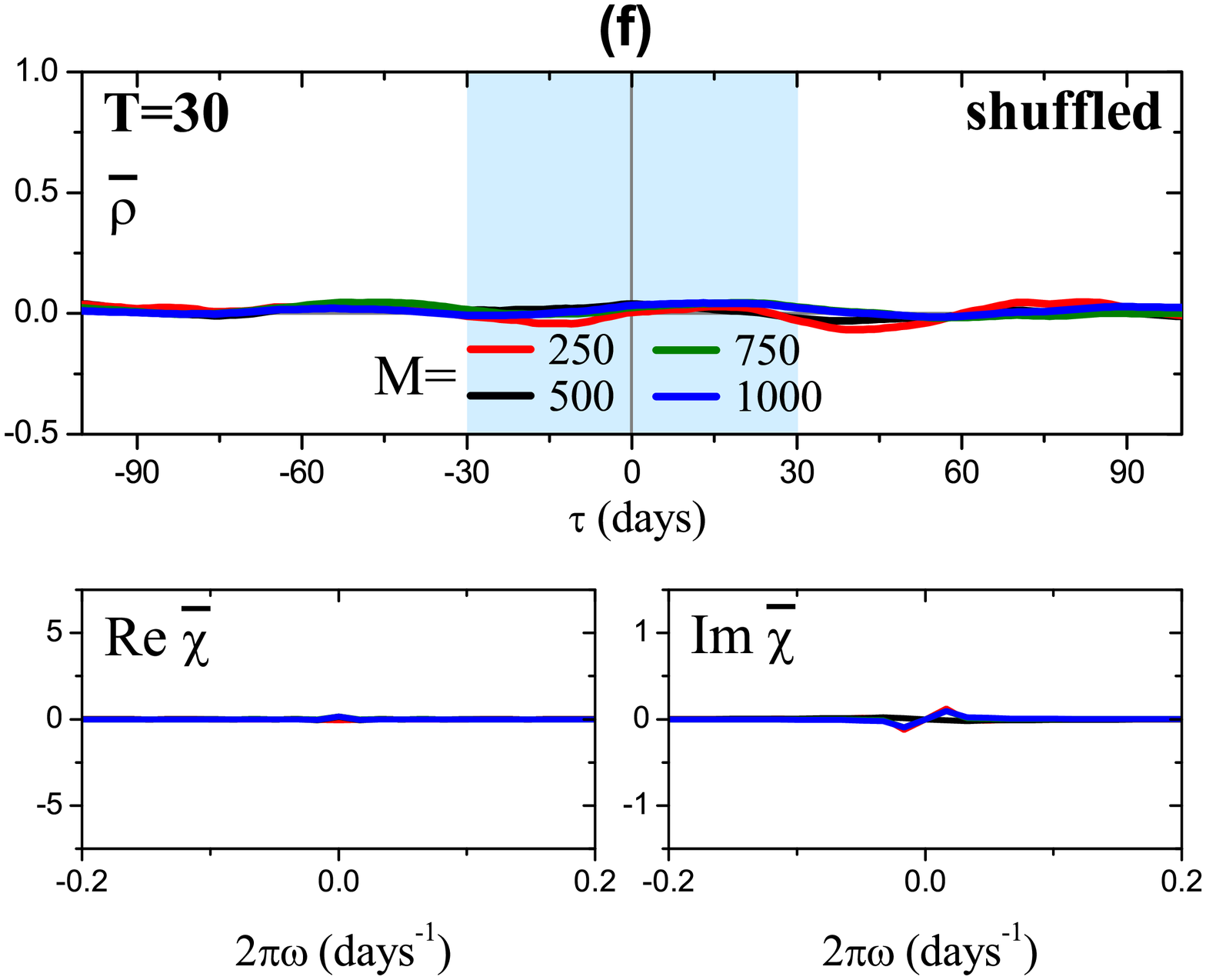}
\end{center}
 \caption{
{\bf Average cross-correlation functions and corresponding susceptibilities calculated using different SMA window sizes for volatility ($T$) and cross-correlations ($M$): Jun 15, 2011 (a), (d); Sep 9, 2011 (b), (e); randomly shuffled returns (c), (f).} The susceptibilities are calculated using the discrete Fourier transform for the range of $\pm 30$ days around zero lag (61 days in total), which is highlighted with a blue background. Bigger values of $T/M$ increase spurious correlations (c) due to smoothing effects.
}
\label{fig:window_1}
\end{figure}

We are now in position to determine causal relations between the individual stock and total market risk, applying the formalism from the previous section. With this aim, we analyze $N = 71$ historical stock prices\cite{yahoo} of the largest US companies in terms of market capitalization, members of the S\&P 500 (Table~\ref{tab:companies}). The historical period considered is between 1994 and 2013, roughly corresponding to 4600 trading days. Being interested in the average market dynamics, we consider a mean value of $\rho[\sigma_i,\sigma_\mathrm{m}]$. However, there is a problem of averaging correlation coefficients since their distribution is highly skewed when the value of $\overline{\rho}$ is close to 1 [top panel in Fig.~\ref{fig:rho_distrib}(a)], what makes them nonadditive quantities. In this regard, a number of methods has been proposed to tackle this issue \cite{Hotelling1953, Hawkins1989}. The simplest one is the Fisher transform \cite{Fisher1915}
\begin{equation}
	\begin{array}{lcl}
	z\left\{\rho\right\} = \frac{1}{2}\ln\left[\frac{1+\rho}{1-\rho}\right] := \tanh^{-1}\left(\rho\right),\\
	 z^{-1}\left\{\rho\right\} =	\tanh\left(\rho\right),
	\end{array}
	\label{eq:fisher}
\end{equation}
which makes the distribution of correlation coefficients approximately normal [bottom panel in Fig.~\ref{fig:rho_distrib}(a)]. In this case, the average correlation might be estimated as
\begin{equation}
	\overline{\rho}[\{\sigma_i\},\sigma_\mathrm{m}](t,\tau) = z^{-1}\left\{\frac{1}{N} \sum\limits_{i=1}^{N} z\left\{\rho[\sigma_i,\sigma_\mathrm{m}](t,\tau)\right\}\right\}
	\label{eq:average_corr}
\end{equation}
with a confidence interval (CI)
\begin{equation}
z^{-1}\left\{z\left\{\overline{\rho}\right\}\pm\frac{z_\mathrm{table}}{\sqrt{N-3}}\right\},
	\label{eq:corr_CI}
\end{equation}
where $z_\mathrm{table}=1.96$ corresponding to the $95\%$ confidence level is further used. When $\overline{\rho}$ is small, the distribution is not skewed and the Fisher transform does not affect it ($z\left\{\rho\right\}\approx\rho$ for small $\rho$) [Fig.~\ref{fig:rho_distrib}(b),(c)]. This average function is subsequently Fourier transformed to obtain the average susceptibility $\overline{\chi}$ using the discrete analogue of Eq.~\ref{eq:susceptibility} for the interval $\tau \in [-\tau^\mathrm{max}, \tau^\mathrm{max}]$. It is also worth noting that the use of an SMA for the calculation of volatilities ($\sigma_i$ and $\sigma_\mathrm{m}$) imposes smoothing on the corresponding time series. Thus, a bigger window of size $M>T$ for the calculation of $\rho$ in Eq.~(\ref{eq:corr}) should be used to avoid spurious correlations [Fig.~\ref{fig:window_1}(c),(f)]. Additionally, Fig.~\ref{fig:rho_distrib}(c) suggests that the averaging over a big number of stocks effectively reduces related undesirable effects.

\begin{figure}[!ht]
\begin{center}
\includegraphics[width=3in]{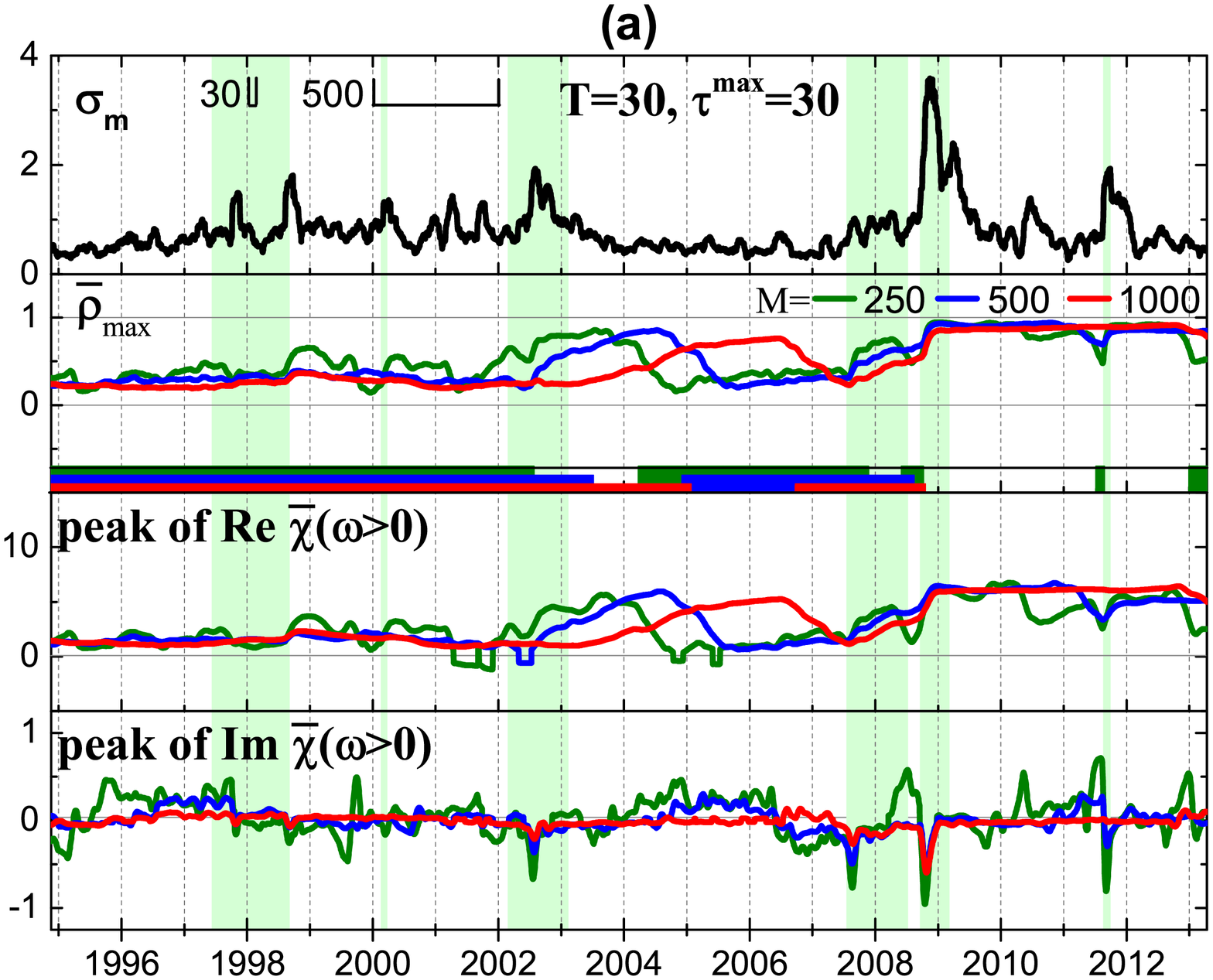}
\hspace{5mm}
\includegraphics[width=3in]{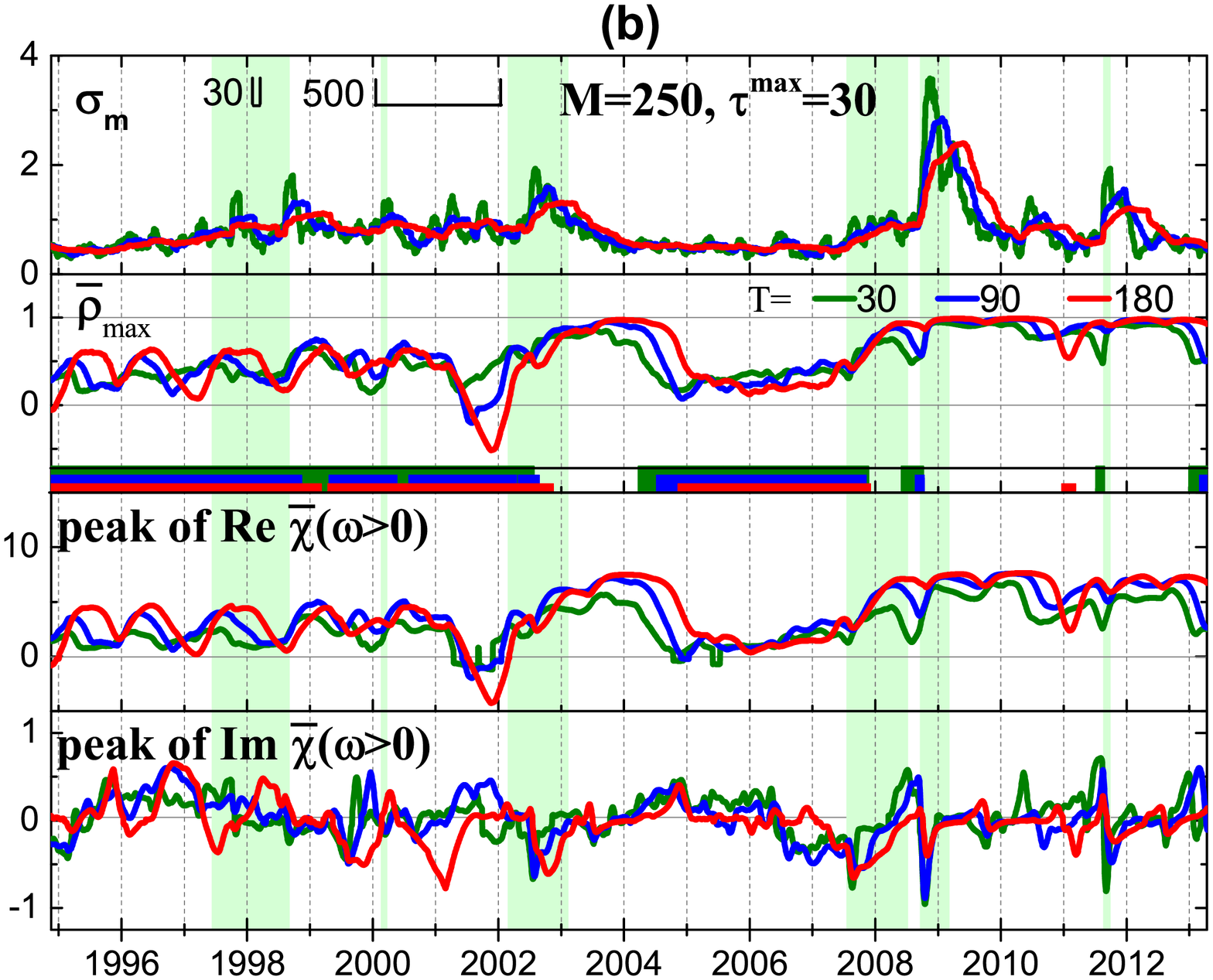}
\end{center}
 \caption{
{\bf Historical dynamics of the (top-bottom) market volatility $\sigma_\mathrm{m}$, maximum value of the average cross-correlation $\overline{\rho}^\mathrm{max}$, peak value of the real and imaginary parts of the average susceptibility $\overline{\chi}(\omega>0)$.} The historical dynamics is calculated for the different SMA windows: $T=30$, $M=250,500,1000$ (a) and $M=250$, $T=30, 90, 180$ (b) days. Filled areas under the $\rho^\mathrm{max}$ panel mark the periods where $\overline{\rho}^\mathrm{max}$ is not significantly bigger than $0.5$. The distance between two labeled dates is 500 trading days and the highlighted periods correspond to the major financial crises described in Fig.~\ref{fig:crashes}.
}
\label{fig:crashes_and_suscept}
\end{figure}

The task at hand requires the series in question to be correlated. For this purpose, we calculate the maximum value of the correlation between the market risk and individual stock risk, $\overline{\rho}^\mathrm{max}$, within the considered range of lag $\pm\tau^\mathrm{max}$. The historical dynamics of this maximum value (second panel in Fig.~\ref{fig:crashes_and_suscept}) suggests that it becomes significantly bigger than $0.5$ near major financial crashes, while in other times the series seem to be weakly correlated. In this respect, one can highlight the US market downturn of 2002 and approximately the 5-year period from the US housing bubble in 2007 until 2013, when almost the linear relationship was observed. For such highly correlated risks, it is feasible to perform causal analysis within the linear response approximation. 

\begin{figure}[!ht]
\begin{center}
\includegraphics[width=3in]{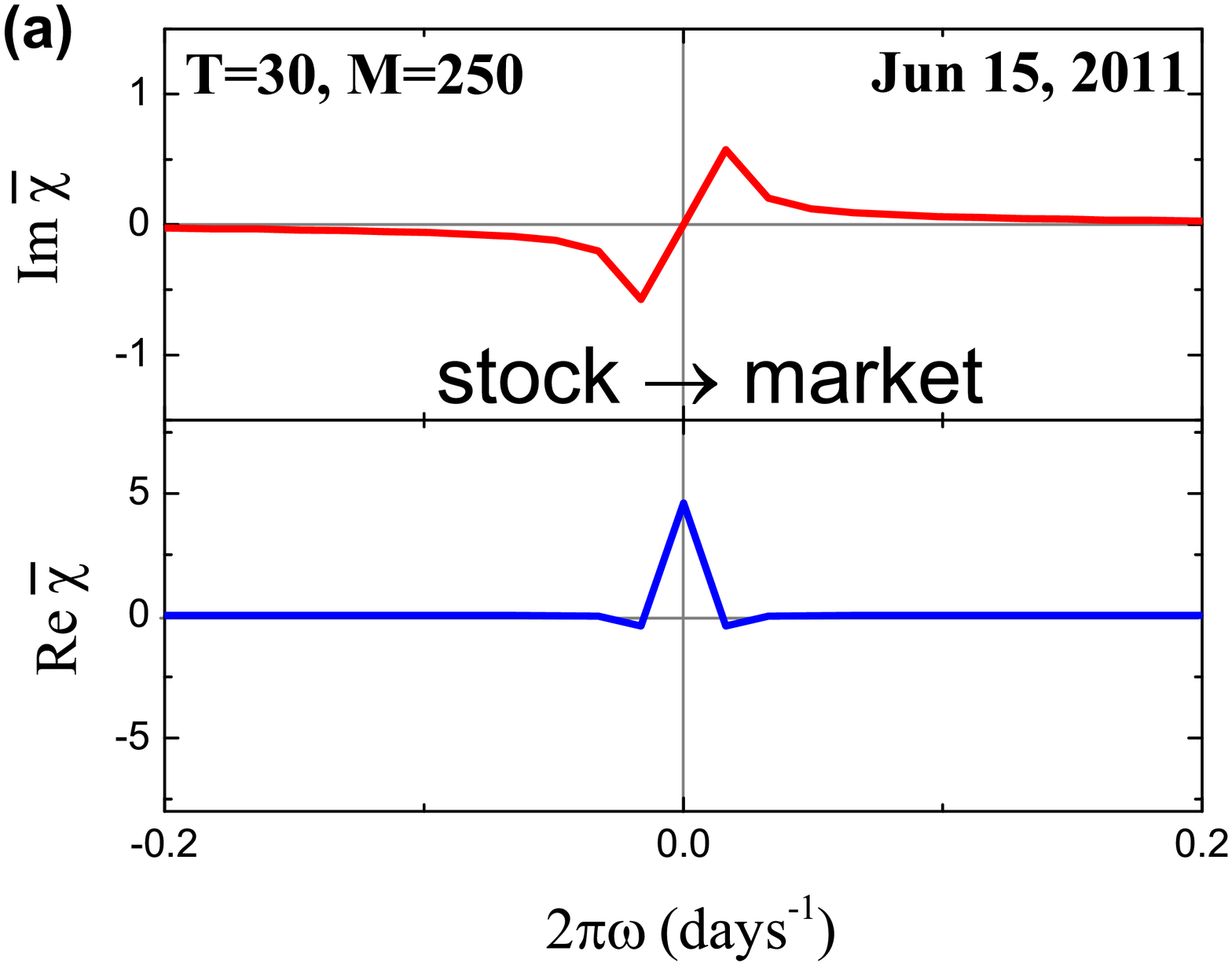}
\hspace{5mm}
\includegraphics[width=3in]{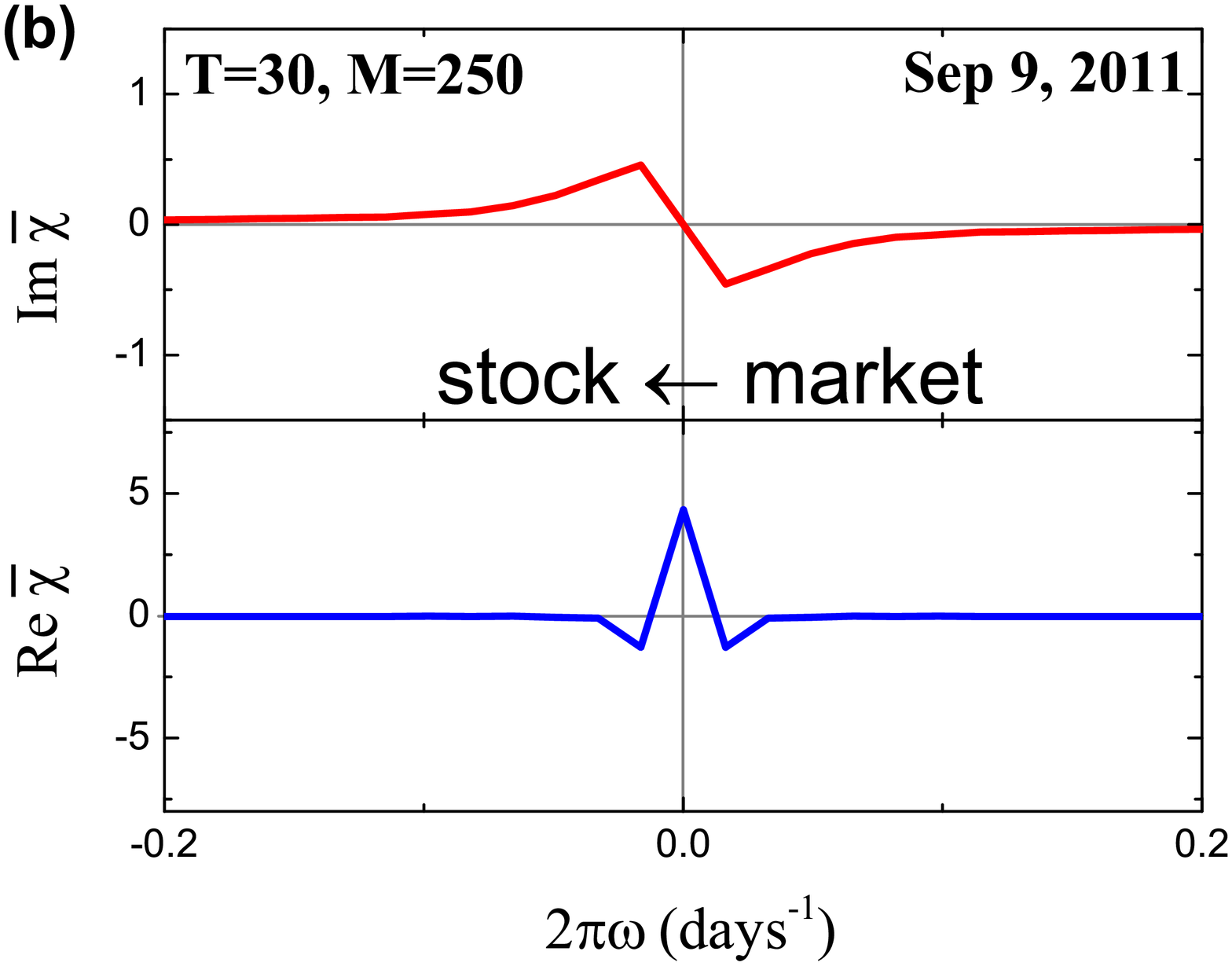}
\end{center}
 \caption{
{\bf Susceptibilities $\overline{\chi}$ for the averaged cross-correlation functions depicted in Fig.~\ref{fig:vol_vs_vol_m_cross_corr}.} The peaks of the imaginary parts hint at the causal relationships between the individual and collective risks: (a) individual stock risks on average tend to influence overall market risk; (b) market risk tends to influence risks of separate stocks. The susceptibilities are calculated using the discrete Fourier transform for the range of $\pm 30$ days around zero lag (61 days in total), which is highlighted with a blue background in Fig.~\ref{fig:vol_vs_vol_m_cross_corr}.
}
\label{fig:vol_vs_vol_m_suscept}
\end{figure}

\begin{figure}[!ht]
\begin{center}
\includegraphics[width=4in]{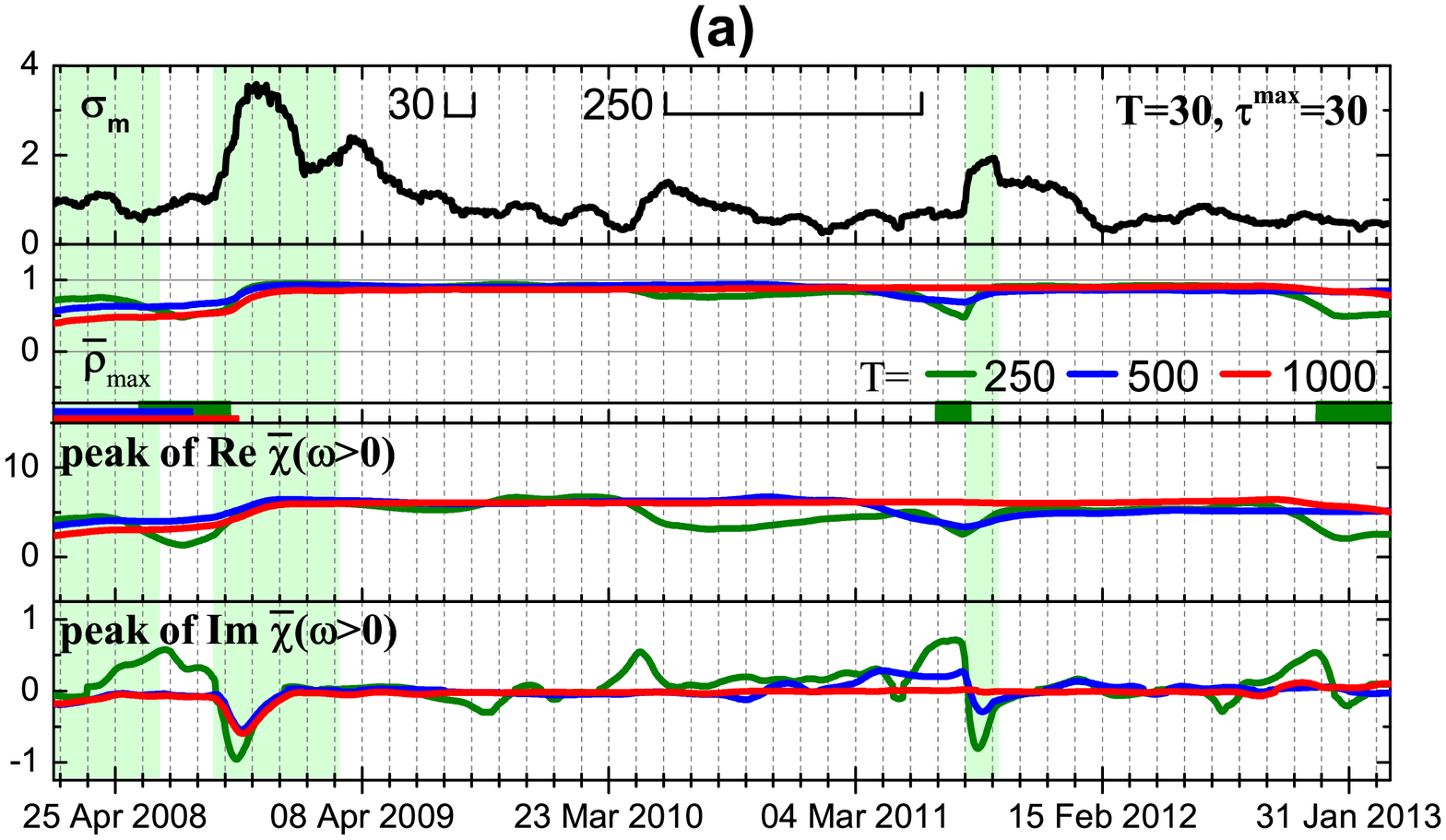}\\
\hspace{0mm}
\includegraphics[width=4in]{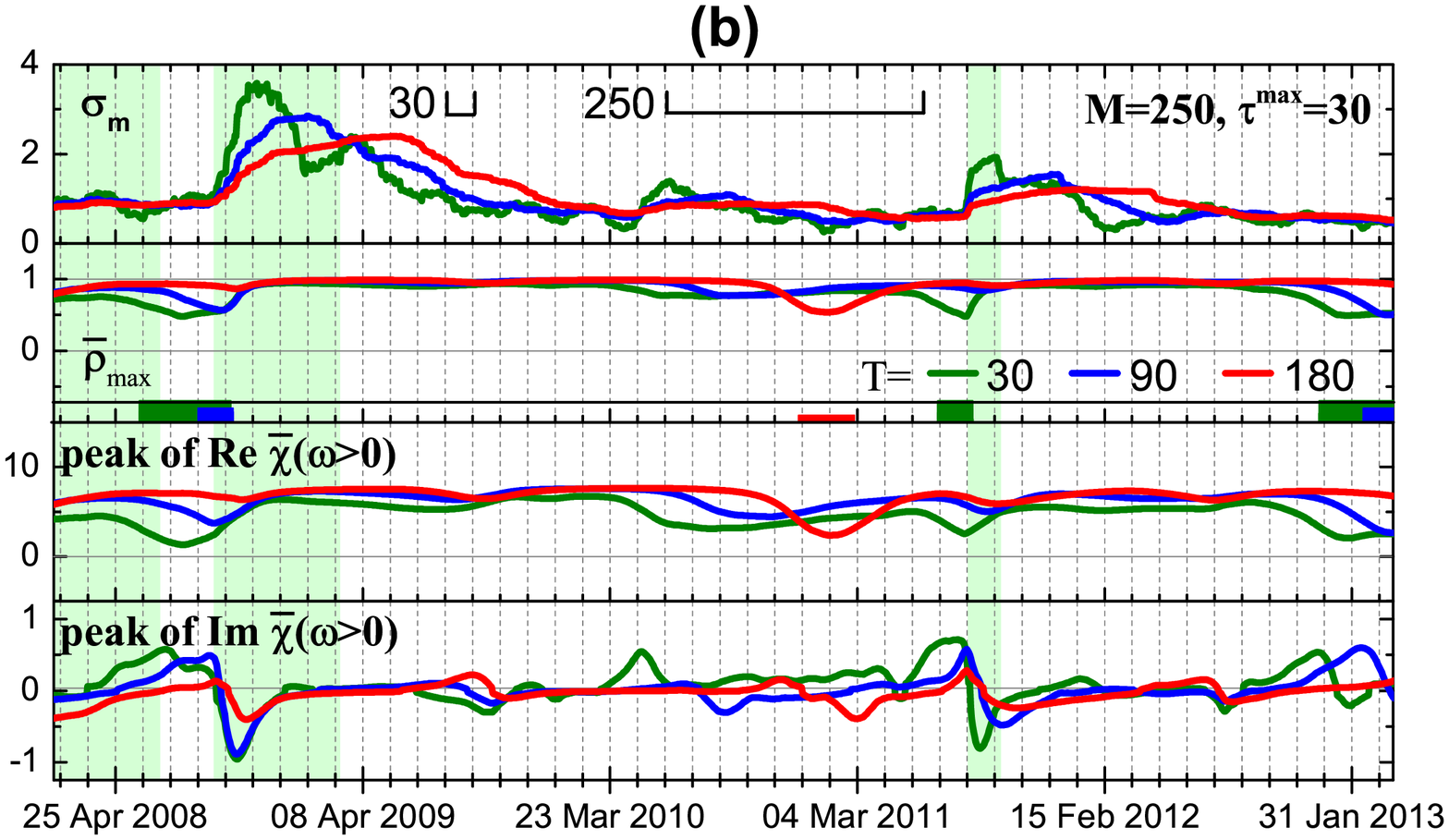}\\
\hspace{0mm}
\includegraphics[width=4in]{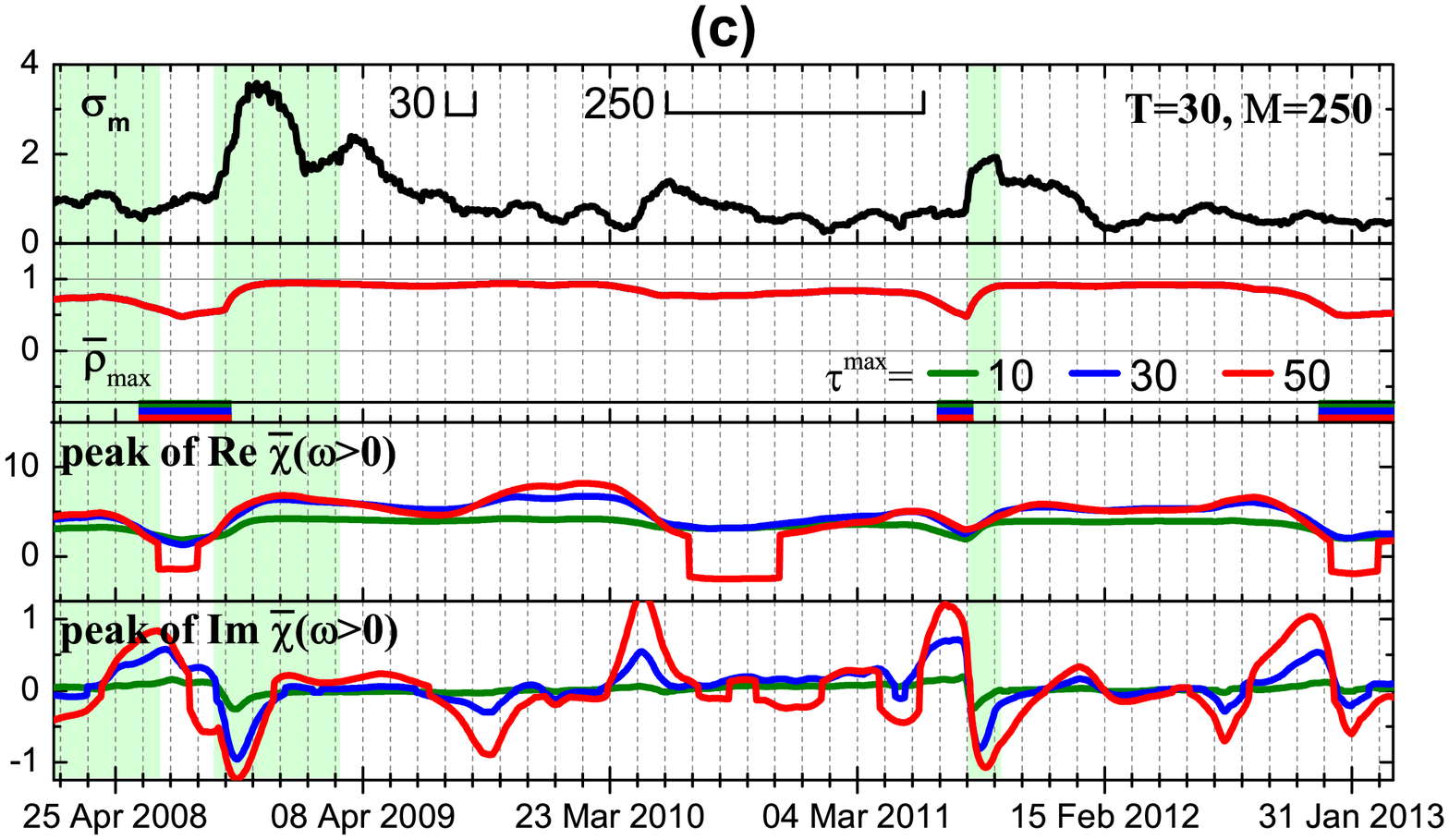}\\
\end{center}
 \caption{
{\bf The same as Fig.~\ref{fig:crashes_and_suscept} but zoomed in for the historical period of the high correlation between individual and systemic risk for different SMA window sizes used for calculation of the volatilities $T$ (a), cross-correlations $M$ (b) and the range of the Fourier transform $\tau^\mathrm{max}$ used for calculation of the average susceptibility (c).}
}
\label{fig:window_2}
\end{figure}

As was mentioned before, typical shapes of $\overline{\rho}$ and $\overline{\chi}$ are depicted in Fig.~\ref{fig:vol_vs_vol_m_cross_corr} and Fig.~\ref{fig:vol_vs_vol_m_suscept} respectively. For instance, causality analysis of these two dates near European sovereign debt crisis reveals that for Jun 15, 2011 [Fig.~\ref{fig:vol_vs_vol_m_cross_corr}(a)] the maximum value of the cross-correlation function is shifted left with respect to zero lag, which is reflected as a negative peak of the imaginary part of the susceptibility for positive frequencies [Fig.~\ref{fig:vol_vs_vol_m_suscept}(a)]. Following the discussion from the previous section, this feature corresponds to the leading influence of individual stock risks on the total market risk. While the opposite situation is observed on Sep 9, 2011 [Fig.~\ref{fig:vol_vs_vol_m_cross_corr}(b) and Fig.~\ref{fig:vol_vs_vol_m_suscept}(a)]. The historical analysis of the average susceptibility dynamics (two bottom panels in Fig.~\ref{fig:crashes_and_suscept}) for the periods with high value of $\overline{\rho}^\mathrm{max}$ reveals two peculiarities. The first one is related to the fact that individual stock risks follow market risk after big crashes. This feature can be viewed as a consequence of herding behavior, when stock risks are trying to reach new equilibrium with overall market risk as a benchmark. This fact is also in agreement with the studies on asymmetric phenomena \cite{Bekaert01012000,asymmetry1,JOFI:JOFI340}, which have shown enhance of volatility and correlations in a bear market. The second peculiarity can be observed, for example, before the Lehman Brothers collapse in 2008 and the European sovereign debt crisis in 2012, when individual stock risks on average start to influence market risk shortly before a crash, while at the crash the direction of influence is reversed. Finally, Fig.~\ref{fig:window_2} shows that this behavior is observed for different window sizes $T$ and $M$, however, use of bigger values of $M$ smooths described effects.

%-------------------------------------------------------------------------------
\section*{Discussion}
We have studied average lead-lag relationships between individual stock and collective market risk in the US stock market using cross-correlation analysis. Our calculations have shown that stock and market volatility are tightly correlated during the periods of financial instability. Furthermore, the correlation functions often possess asymmetries with respect to zero lag, which is a potential sign of a causal dependence between the risks within the linear response approximation. Having analyzed historical data for 1994--2013, we have found similar patterns near the last major crashes. Firstly, after a financial crash individual stock risks tend to follow collective market behavior. Secondly, the opposite influence is observed when stock risks on average start to influence market risk before particular crashes, for instance, the Lehman Brothers collapse in 2008 or the European sovereign debt crisis in 2012. Eventual market adjustment after the crash leads to the restoration of a symmetric shape of the average cross-correlation function and decrease of its maximum value. This is also reflected in the Fourier transform of the cross-correlation known as susceptibility. For this complex function, reversal of the causal dependence corresponds to the change of the sign of its imaginary part, while the real part remains unaffected, and the absence of the dependence results in zero value of the imaginary part. We suggest that the observed patterns might be interpreted as a manifestation of herding behavior, when economic performance of separate companies systematically does not meet expectations of investors, creating the panic across the market. Wherein after the crash, financial risks of separate companies adapt to a new reality with overall market performance as a psychological benchmark.

% You may title this section "Methods" or "Models". 
% "Models" is not a valid title for PLoS ONE authors. However, PLoS ONE
% authors may use "Analysis" 
%\section*{Materials and Methods}

% Do NOT remove this, even if you are not including acknowledgments
\section*{Acknowledgments}
We are grateful to L.~Pietronero, Y.~Roudi, J.~Suorsa, J.~Edge and anonymous referees for providing useful comments and discussion. This work is supported by Nordita and VR VCB 621-2012-2983.

%\section*{References}
% The bibtex filename
\bibliography{stockscorr}

\section*{}
%\begin{figure}[!ht]
%\begin{center}
%%\includegraphics[width=4in]{figure_name.2.eps}
%\end{center}
%\caption{
%{\bf Bold the first sentence.}  Rest of figure 2  caption.  Caption 
%should be left justified, as specified by the options to the caption 
%package.
%}
%\label{Figure_label}
%\end{figure}

\section*{}
%\begin{table}[!ht]
%\caption{
%\bf{Table title}}
%\begin{tabular}{|c|c|c|}
%table information
%\end{tabular}
%\begin{flushleft}Table caption
%\end{flushleft}
%\label{tab:label}
% \end{table}

\begin{table}[!ht]
\caption{
\bf{List of the companies which stock prices are used for the calculations in the paper.}}
\footnotesize
\begin{tabular}{lll|lll}
\hline
\hline
\noalign{\smallskip}
Ticker & Name & Sector & Ticker & Name & Sector \\
\noalign{\smallskip}
\hline
\noalign{\smallskip}
ABT & Abbott Laboratories & Hea &
AIG & American International Group, Inc. & Fin \\
AMGN & Amgen Inc. & Hea &
APA & Apache Corp. & Bas \\
APC & Anadarko Petroleum Corp. & Bas &
AAPL & Apple Inc. & Con \\
AXP & American Express Company & Fin &
BA & The Boeing Company & Ind \\
BAC & Bank of America Corp. & Fin &
BAX & Baxter International Inc. & Hea \\

BMY & Bristol-Myers Squibb Company & Hea &
C & Citigroup, Inc. & Fin \\
CAT & Caterpillar Inc. & Ind &
CELG & Celgene Corporation & Hea \\
CL & Colgate-Palmolive Co. & Con &
CMCSA & Comcast Corporation & Ser \\
COP & ConocoPhillips & Bas &
COST & Costco Wholesale Corp. & Ser \\
CSCO & Cisco Systems, Inc. & Tec &
CVS & CVS Caremark Corp. & Ser \\

CVX & Chevron Corp. & Bas &
DD & E. I. du Pont de Nemours and Co. & Bas \\
DE & Deere \& Company & Ind &
DELL & Dell Inc. & Tec \\
DHR & Danaher Corp. & Ind &
DIS & The Walt Disney Company & Ser \\
DOW & The Dow Chemical Company & Bas &
EMC & EMC Corporation & Tec \\
EMR & Emerson Electric Co. & Tec &
EOG & EOG Resources, Inc. & Bas \\

EXC & Exelon Corp. & Uti &
F & Ford Motor Co. & Con \\
GE & General Electric Company & Ind &
HAL & Halliburton Company & Bas \\
HD & The Home Depot, Inc. & Ser &
HON & Honeywell International Inc. & Ind \\
HPQ & Hewlett-Packard Company & Tec &
IBM & International Business Machines Corp. & Tec \\
INTC & Intel Corp. & Tec &
JNJ & Johnson \& Johnson & Hea \\

JPM & JPMorgan Chase \& Co. & Fin &
KO & The Coca-Cola Company & Con \\
LLY & Eli Lilly and Company & Hea &
LOW & Lowe's Companies Inc. & Ser \\
MCD & McDonald's Corp. & Ser &
MDT & Medtronic, Inc. & Hea \\
MMM & 3M Company & Cng &
MO & Altria Group Inc. & Con \\
MRK & Merck \& Co. Inc. & Hea &
MSFT & Microsoft Corp. & Tec \\

NKE & Nike, Inc. & Con &
ORCL & Oracle Corporation & Tec \\
OXY & Occidental Petroleum Corp. & Bas &
PEP & Pepsico, Inc. & Con \\
PFE & Pfizer Inc. & Hea &
PG & The Procter \& Gamble Company & Con \\
PNC & The PNC Financial Services Group & Fin &
SLB & Schlumberger Limited & Bas \\
SO & Southern Company & Uti &
T & AT\&T, Inc. & Tec \\

TGT & Target Corp. & Ser &
TJX & The TJX Companies, Inc. & Ser \\
TXN & Texas Instruments Inc. & Tec &
UNH & UnitedHealth Group Incorporated & Hea \\
UNP & Union Pacific Corp. & Ser &
USB & U.S. Bancorp & Fin \\
UTX & United Technologies Corp. & Ind &
VZ & Verizon Communications Inc. & Tec \\
WFC & Wells Fargo \& Company & Fin &
WMT & Wal-Mart Stores Inc. & Ser \\

XOM & Exxon Mobil Corp. & Bas &
& & \\
\noalign{\smallskip}
\hline
\hline
\end{tabular}
\begin{flushleft}
 Sectors are defined as basic materials (Bas), conglomerate (Cng), consumer goods (Con), financial (Fin), healthcare (Hea), industrial goods (Ind), services (Ser), technology (Tec) and utilities (Uti).
\end{flushleft}
\label{tab:companies}
\end{table}

\end{document}